# Frequency Tunable Magnetostatic Wave Filters With Zero Static Power Magnetic Biasing Circuitry

Xingyu Du[1], Mohamad Hossein Idjadi[1], Yixiao Ding[1], Tao Zhang[1], Alexander J. Geers[1], Shun Yao[1], Jun Beom Pyo[1], Firooz Aflatouni[1], Mark Allen[1], and Roy H. Olsson III[1*]

1. Department of Electrical and Systems Engineering, University of Pennsylvania, Philadelphia, PA, USA

Correspondance: Roy H. Olsson III (rolsson@seas.upenn.edu)

**Abstract**

A single tunable filter simplifies complexity, reduces insertion loss, and minimizes size compared to frequency switchable filter banks commonly used for radio frequency (RF) band selection. Magnetostatic wave (MSW) filters stand out for their wide, continuous frequency tuning and high-quality factor. However, MSW filters employing electromagnets for tuning consume excessive power and space, unsuitable for consumer wireless applications. Here, we demonstrate miniature and high selectivity MSW tunable filters with zero static power consumption, occupying less than 2 cc. The center frequency is continuously tunable from 3.4 GHz to 11.1 GHz via current pulses of sub-millisecond duration applied to a small and nonvolatile magnetic bias assembly. This assembly is limited in the area over which it can achieve a large and uniform magnetic field, necessitating filters realized from small resonant cavities micromachined in thin films of Yttrium Iron Garnet. Filter insertion loss of 3.2 dB to 5.1 dB and out-of-band third order input intercept point greater than 41 dBm are achieved. The filter's broad frequency range, compact size, low insertion loss, high out-of-band linearity, and zero static power consumption are essential for protecting RF transceivers and antennas from interference, thus facilitating their use in mobile applications like IoT and 6G networks.

The growth of multi-band and high frequency communication systems has resulted in single bandpass filter technologies being unable to satisfy the filtering requirements for all bands due to the crowded radio-frequency (RF) spectrum.[1, 2, 3] This is especially problematic as frequencies are scaled beyond the spectrum allocated for 5G (3-6 GHz), where RF silicon-on-insulator switches exhibit unacceptably high loss when utilized in switched filter banks.[4, 5] For example, the FR3 band from 7.125 to 24.25 GHz under exploration for 6G networks will require extensive innovations in both RF switch[4, 6] and acoustic filter [7, 8, 9, 10] technologies if it is to adopt the massively parallel switched acoustic filter banks utilized in 4G and 5G networks. As compared to the traditional implementation of a switched filter-bank, a single tunable filter has great potential to reduce the system cost, size, complexity, and remove entirely the additional switch paths loss.[2, 11, 12, 13] Tunable bandpass filters are needed in applications beyond 6G wireless such as cognitive radios[14], frequency hopped receivers[15], satellite communications[16], base stations[17], and multiband radar[18].

For the frequency spanning from S band to X band (2-12 GHz), numerous tunable filters have been developed to eliminate out-of-band noise while preserving the in-band signal. However, most of these filters have a limited frequency tuning range. This is also elaborated on in Supplementary Note 1. Mechanically tunable filters offer high quality factor (Q) but require external circuitry and motors for tuning and have a limited center frequency range with a maximum center frequency tuning ratio of 1.2:1[19, 20, 21]. RF MEMS enabled tunable electromagnetic cavity filters achieve a center frequency tuning ratio up to 2:1 and high-power handling capabilities but are relatively large and sensitive to shock and vibration [22, 23, 24]. Varactor based tunable filters are small, have fast tuning speeds, and moderate center frequency tuning ratio of 2:1, but suffer from low Q values[25, 26, 27, 28]. To the best of our knowledge, the highest reported $S_{12}$ quality factor (Q-factor) is about 46 at L band and is heavily dependent on frequency[26]. This limits the minimum filter bandwidth and the steepness of the filter skirts. In addition, while a tunable filter attenuates out-of-band interferer signals, the intermodulation distortion of the filter could permit out-of-band signals a path to mix into the filter passband which can deteriorate receiver signal-to-noise ratio. Although much higher linearity can be achieved for mechanical and RF MEMS based tunable filters, the out-of-band (OOB) 3rd order input referred intermodulation intercept point (IIP3) of other tunable filters is usually in the range of 23.5 dBm[29] to 27 dBm[30].

Tunable filters realized using magnetostatic wave resonators (MSWR) are a promising technology to fulfill the demands of broad and continuous frequency tuning range with high quality factor >1000[31]. Magnetostatic waves (MSW), also known as dipolar spin waves, are long wavelength spin waves, where the magnetic dipolar interactions dominate both electric and exchange interactions.[32] Because the MSW group velocities are slower than that of electromagnetic waves and are variable with applied bias magnetic field, magnetic field tunable magnetostatic wave filters (MSWF) with a wide frequency tuning range are possible.[32, 33] Micrometer thick, single crystal yttrium iron garnet (YIG) thin films exhibit the lowest damping for MSW and thus the smallest propagation loss as compared to other common ferromagnetic materials.[34] As a result, previous studies have demonstrated YIG based MSWR with large quality factors of 3600 at 9 GHz[35] and 5259 at 4.77 GHz[31].

Despite the small size and high Q of YIG MSWRs, YIG tunable MSWFs still suffer from large size, high power consumption, and slow tuning speed from the use of bulky and energy intensive electromagnets to supply the necessary magnetic bias field for MSWFs.[36] SI Note 1 shows a size comparison between this study and two leading commercial YIG-based tunable filters. These filters, incorporating electromagnet drivers for YIG sphere resonator frequency tuning, result in sizes exceeding 23 cm$^3$ and power consumption surpassing 2 W. These limitations hinder their applicability in Internet of Things (IoT) and mobile phone technologies.

Filters based on YIG sphere resonators have demonstrated low loss across a broad tuning frequency range (9:1). YIG sphere resonators, however, are too large to fit within the miniature magnetic bias circuit reported here. [37, 38] Previously reported planar geometry MSWR formed through standard microfabrication processes have a form factor compatible with the reported small, tunable, magnetic bias circuits. Previous filters realized from MSWR, however, exhibited 20 to 32 dB insertion loss when operating with a wide frequency tuning range between 2 to 12 GHz.[39] This is mainly due to the difficulty of obtaining large coupling and a well matched impedance across a broad frequency range. Low insertion loss planar YIG tunable filters were only demonstrated over a limited frequency range: ~5.3 dB loss with a center frequency tuning ratio of 1.5:1 in X band[40] and 5.8-6.4 dB loss with a tuning ratio of 1.5:1 in X and $K_u$ bands [41]. In order to meet the requirements of insertion loss and out-of-band suppression, a planar MSWR with a large area of 4 mm × 10 mm[42] or a five layer stack MSWR with dimensions of 2 mm × 2 mm × 0.62 mm were needed[41].

In this study, we demonstrate miniature, narrowband, frequency tunable filters (3.3:1) with zero static power consumption and exceptional out-of-band linearity. Fig. 1 a-c depicts the device assembly and shows images of the tunable filter assembly with a total volume of only 1.68 cm$^3$. To tune the cavity center frequency, current pulses were applied to AlNiCo pieces in the magnetic bias assembly to alter their nonvolatile magnetic remanence. Using this approach, the tunable magnetic bias circuit only consumes transient power to tune the magnetic field and filter frequency and enables the frequency tunable filter to operate without any steady state power consumption. Magnetostatic surface waves were utilized, where an in-plane magnetic bias field is established in the YIG perpendicular to the direction of magnetostatic wave propagation. The MSWF is based on cavities microfabricated in a YIG thin film with straight edge reflectors. By using planar microfabricated YIG cavities to form MSWFs, the size remains small such that the filters can fit within in the small, tunable bias assembly, which can only produce large and uniform magnetic fields over a small area. To optimize for low insertion loss, aluminum input and output transducers were placed directly on the YIG film to efficiently excite and collect the magnetostatic waves with high coupling. The geometrical parameters were also optimized based on the equivalent circuit shown in Fig. 1d to enable impedance matching to 50 Ω over the broad 3.3:1 frequency tuning range. These innovations enabled low filter insertion loss of 3.2-5.1 dB across the entire 3.4-11.1 GHz frequency tuning range with a YIG filter occupying only 200x70 µm$^2$ of area.

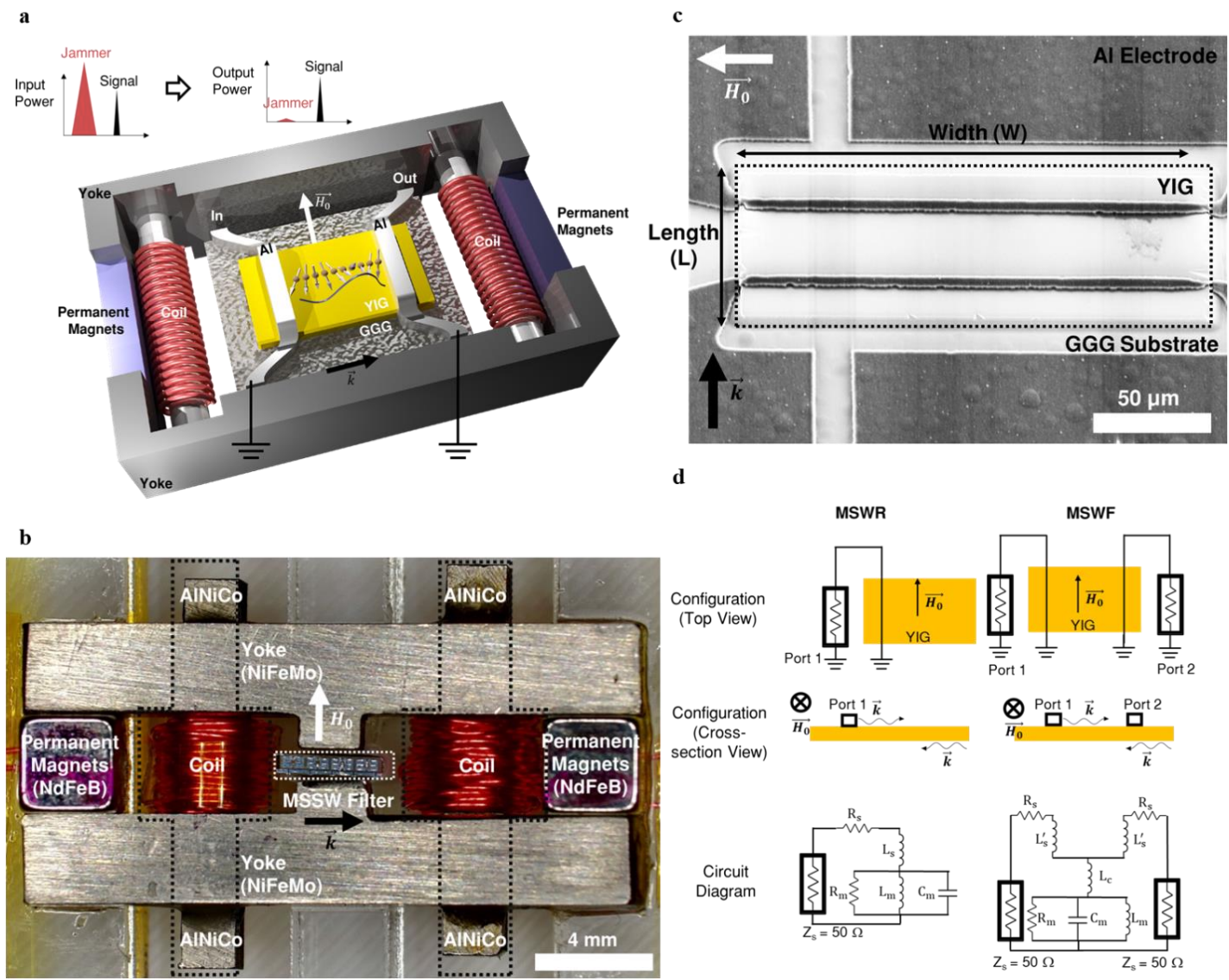

**Fig. 1 Tunable Bandpass Filter with Magnetic Biasing Circuit.** The MSWF were placed in the center of the magnetic biasing component. The aluminum transducers were placed on top of the YIG cavity. The magnetic biasing component consists of two permanent magnets, two shunt magnets wrapped with coils, and two magnetically permeable yokes which concentrate the magnetic flux in the MSWF. **a,** Reconfigurable MSWF concept. **b,** Optical microscope image of the fabricated device assembly. **c,** Scanning Electron Microscope image showing the aluminum transducers on top of the YIG cavity. This device has a width (W) of 200 μm and length (L) of 70 μm. **d,** Summary of device schematic diagrams and equivalent single-mode circuit models.

**Magnetostatic Wave Resonator (MSWR):**

Straight edge MSWR consisting of a ferrimagnetic resonant cavity made of a 3.3 μm film of YIG grown on top of a Gadolinium Gallium Garnet (GGG) substrate were patterned into a rectangular shape by wet etching. The transducers made of 2 μm thick aluminum (Al) microstrips were fabricated on top of the YIG. The transducers width is approximately 7 μm unless stated otherwise. Fig. 1c shows one typical fabricated device, where the width (W) is defined as the coupling length of the Al electromagnetic transducer on the MSW, whereas the length (L) is defined as the MSW cavity length in the direction of MSSW propagation. The fabricated MSWR was first measured under the magnetic probe station where two electromagnets were used to generate the magnetic bias field, as further illustrated in Supplementary Notes 2 and 3.

Inside the YIG cavity, the MSW is stimulated by inductive antennas, inducing oscillating magnetic fields through RF currents, as shown in Fig. 1d. The structure of the YIG cavity consists of two parallel reflecting interfaces formed by wet-etched YIG edges. As a result, spin waves entering the YIG cavity circulate coherently with minimal damping. By placing a single Al transducer on top of the YIG, the YIG cavity is configured as a MSWR. Alternatively, the MSWR employing two Al transducers produces a filter-like bandpass frequency response with high out-of-band rejection. The cross-section view depicts the unidirectional propagation of the MSW, which exclusively propagate along the surface of the YIG and are reflected onto the other surface.[43, 44] As a result, the propagation path is not reciprocal between the two ports. This MSWR response can be represented by an equivalent circuit model of a parallel RLC circuit in series with an ohmic series resistance ($R_s$) and a self-inductance ($L_s$).[31] The resonance tank within the MSWR exhibits maximum impedance at the resonance frequency. Consequently, the return loss displays a dip at the resonance frequency. Two-port MSWF can be modeled by connecting the resonance tank with a coupling inductor $L_c$, which considers the direct electromagnetic coupling between the two ports where the series inductance satisfies the relationship $L_s = L'_s + L_c$.

A single RLC tank circuit with magnetostatic resistor $R_m$, magnetostatic inductor $L_m$, magnetostatic capacitor $C_m$ only captures the response at one frequency. The complete impedance response can be modeled by the multi-mode circuit model in Fig. 2a with the number of modes, p. Fig. 2b compares the modeled and measured MSWR input impedance of a MSWR with W = 200 μm and L = 70 μm with an Al transducer width of 4 μm. The impedance of the resonance frequency, $f_s$, is typically evident without circuit modeling. However, the single-mode circuit model plays a crucial role in accurately determining the precise frequency and magnitude of the anti-resonance frequency, $f_p$. At this point, the impedance of the fundamental mode becomes extremely small, and the influence of other modes can impede its observation. Moreover, the multi-mode circuit successfully predicts the impedance response across a wide frequency range. Supplementary Note 4 details the circuit modeling procedure. The circuit modeling and analysis were performed directly on the measured data without any de-embedding process.

The single-mode circuit model has proven to be effective in accurately predicting bandwidth, magnitude, and phase of the impedance around the peak frequency of the MSWR. A MSWF's performance can be optimized by increasing the coupling coefficient ($K^2$), Q, and figure of merit (FoM = $K^2Q$) of the MSWR mode. Thus, it has been utilized to study these important performance parameters and the magnetostatic resistance, $R_m$. Fig. 3 compares the influence of the width of the MSWR on these parameters, with a fixed MSWR length of 70 μm. Increasing the width of the YIG cavity results in a higher $K^2$, FOM, and magnetostatic resistance, $R_m$, as the coupling distance increases. A maximum $K^2$ = 2.4 % is achieved at a width of 600 μm. The Q-factor does not show a significant change with the width of the YIG cavity. However, the Q-factor generally increases with frequency. This helps to achieve an almost constant filter bandwidth with center frequency tuning, which is one of the advantages of MSSW filters to achieve constant data rates at various frequencies. The maximum Q-factor measured is 1313, which is for the MSWR with width of 150 μm at a frequency of 11.6 GHz. The increase in Q with frequency can be attributed to the increase in $R_m$ with frequency, while the $R_s$ remains constant. This increased energy storage leads to a higher Q value. Similarly, the drop in coupling with frequency is caused by the decrease in $L_m$ with frequency, while the $L_s$ remains constant. Since the $K^2$ is determined by the ratio of $L_m$ to $L_s$, a decrease in $L_m$ leads to a decrease in coupling. This high Q-factor demonstrates the excellent selectivity and efficiency of the MSWR in achieving narrow bandwidths and minimizing signal losses. The plot of magnetostatic resistance, $R_m$, vs. width shows a linearly increasing relationship. This agrees with previous theoretical calculations of the radiation resistance increase with the coupling distance.[45, 46] In a simple microstrip model, the transducer self-inductance and self-resistance should be linearly proportional to its length. However, in practical scenarios, factors such as contact resistance and the resistance and inductance associated with the transducer routing introduce complexities. As a result, the total measured series resistance and inductance do not scale directly with the width of the device. For instance, the total series resistance is approximately 0.84 Ω and 2.2 Ω, and the total series inductance is 0.22 nH and 0.68 nH for W = 150 μm and 600 μm resonators respectively.

The series resistance and inductances increase by 2.6~3.1x, where a 4x scaling is expected. This suggests significant unwanted series components arise from factors such as contact resistance or other electrical routing. Future studies could explore the use of thicker aluminum transducers or new layout designs, which could increase the coupling and quality factor of the devices.

To design a MSWR with better FoM, the width effect of the Al transducers and the length effect of MSWR are also discussed in Supplementary Notes 5 and 6. Impedance matching plays a crucial role in the design of a low loss filter, as further illustrated in Supplementary Note 7. Overall, W= 150-200 μm MSWR are better matched to the 50 Ω source impedance at high frequencies whereas the W = 600 μm MSWR are matched to 50 Ω at low frequency. Although higher $K^2$ and FoM were achieved in the wide MSWR, the impedance mismatch causes the insertion loss for W = 600 μm to be higher than that of W= 100 or 200 μm when taken across the tunable frequency range.

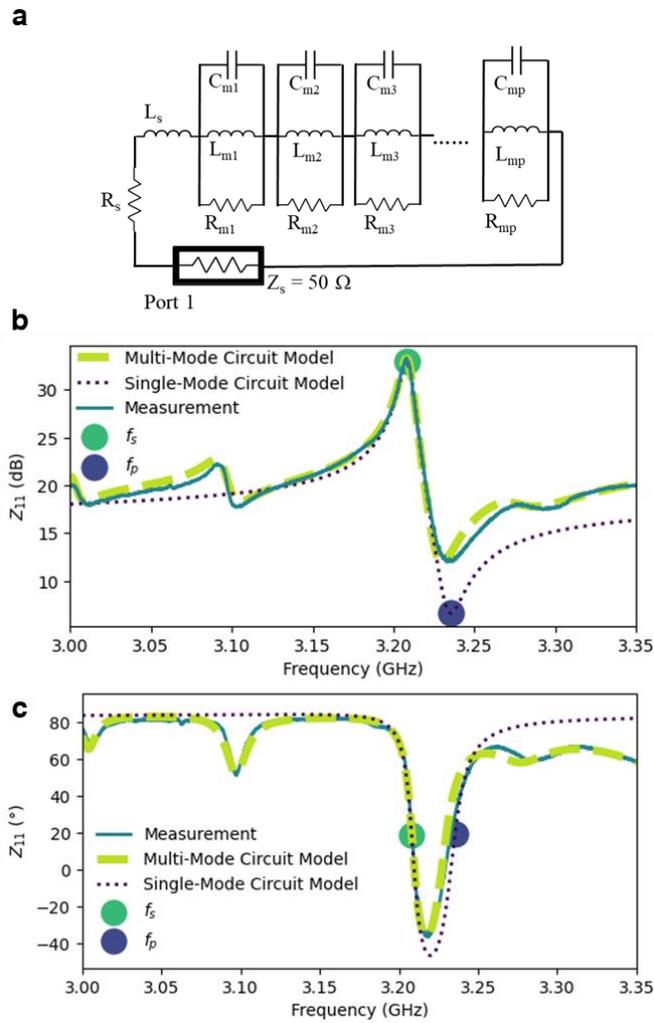

**Fig. 2 Circuit Modeling for MSWR. a,** multi-mode circuit model for MSWR. **b,** comparison of the impedance magnitude of the measured MSWR, single mode circuit model, and multi-mode circuit model. **c,** comparison of impedance phase of the measured MSWR, single mode circuit model, and multi-mode circuit model. The resonance frequency, $f_s$ and anti-resonance frequency, $f_p$, of the single-mode circuit model has been labeled. This MSWR is with W = 200 μm and L = 70 μm with Al transducer width of 4 μm.

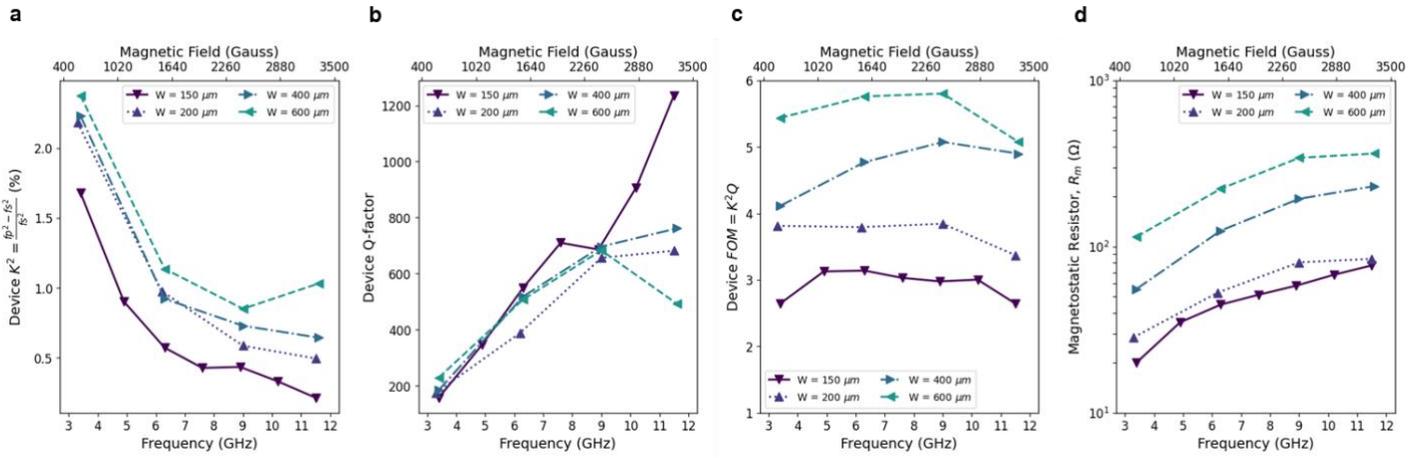

**Fig. 3 Comparison of MSWR for various YIG cavity widths.** The effect of YIG cavity width on **a,** the device coupling coefficient, **b,** the device Q-factor, **c,** device FOM, **d,** the magnetostatic resistance, $R_m$.

**Magnetostatic Wave Filter (MSWF)**

Supplementary Note 8 illustrates the tunability of the MSWF via applied magnetic field. The relationship of the main resonance frequency with respect to the applied external magnetic bias field is linear with a slope of 2.9 MHz/Gauss. Fig. 4 shows the typical $S_{12}$ frequency responses of the MSWF. All these MSSW filters exhibited less than 10 dB insertion loss with greater than 20 dB out-of-band isolation.

As was discussed in the previous section, the FoM increases with increasing width, resulting in a lower insertion loss for the wide MSWF at 3.4 GHz. However, the insertion loss of W = 600 μm at 9.1 GHz is not lower than the W = 200 μm filter. This is because the main resonant tank becomes over-coupled to the source impedance of 50 Ω, leading to significant reflection loss. The length of the MSSW filter, as explained in the supplementary notes, does not have a significant effect on the FoM and $R_m$. Hence, the insertion losses of the L = 70 μm and 140 μm devices are similar. Utilizing wide MSWRs presents a drawback in terms of decreased out-of-band rejection. While wider widths result in increased coupling to the MSW, they also lead to higher direct electromagnetic wave coupling between port one and port two. However, it is worth noting that the direct electromagnetic wave coupling can be mitigated to some extent by increasing the propagation distance of the MSW. This increased propagation distance helps to reduce the direct electromagnetic wave coupling between the ports, thereby improving the out-of-band rejection performance of the MSWF.

Previous studies have shown the dispersion relations for ferrimagnetic films with finite dimensions[47]. The finite film width of ferrimagnetic films introduces a multitude of propagating modes known as width modes. MSW propagate exclusively along the surfaces of the YIG film and undergo reflection to the other surface at the straight edges. This reflection process gives rise to the formation of circulating wave patterns within the film. These patterns result in resonance when the round-trip phase of the MSWs inside the cavity equals $2\pi$. It is important to note that MSWs exhibit a strongly dispersive nature, causing the resonance modes at $2\pi$, $4\pi$, $6\pi$, etc. to be closely spaced in frequency. This contrasts with cavities that utilize low dispersion waves such as electromagnetic and acoustic waves. Further details on the calculation of length and width modes can be found in Supplementary Note 9. Due to the unique dispersion relationship, the W= 600 μm MSWF results in a smaller spacing between two adjacent width modes, due to an increase in the dispersion of the wider device. Additionally, an increase in width results in a frequency shift for the main resonance mode, as observed in the dispersion curve where a constant wavenumber corresponds to a higher frequency as the width increases. This frequency shift is further supported by the measurement results, which demonstrate that narrow MSWF exhibits a lower resonance frequency for the fundamental mode. Furthermore, it is evident from the calculations and measurements that MSW becomes more dispersive at higher frequencies in wider MSWF. A longer YIG cavity showed more spurious responses as more main resonant peaks were observed in the device with L =105 and 140 μm than the device with W = 150 μm, L = 70 μm.

Supplementary Note 11 describes the measurement for power handling capability of the MSWF with W = 150 μm and L = 70 μm. Out-of-band signals that are outside the limits-imposed by the MSW dispersion relationship, are unable to excite MSW waves, and thus are unaffected by increases in input power. In-band $S_{12}$ decreases with the increase of the input power beyond -20 dBm at all the four measured frequencies while the out-of-band response remains unchanged. The in-band, input 1dB compression point is -17 dBm (min) at 3.4 GHz and a -14 dBm (max) at 8.9 GHz. The in-band, output 1dB compression point exhibits a selective limiting effect associated with MSW propagation in YIG.[48] When the input power is above $P_{1dB}$ and below approximately 8 dBm, the insertion loss is increasing, and the output power is saturating with increasing input power. When the input power is above 9 dBm, the direct inductive coupling between the input and output aluminum transducers dominates the insertion loss and the output once again increases linearly with input power. Such self-limiting behavior can be useful to protect receivers from damage under large in-band interference.

Supplementary Note 11 shows the in-band and out-of-band IIP3 measurements for W = 150 μm and L = 70 μm. The in-band IIP3 does not change significantly with the resonance frequency but is a strong function of tone spacing. At a tone spacing of 1 MHz, the IIP3 displays the minimum value of -8 dBm at 7.6 GHz and -11 dBm at 10.1 GHz. The filter 3dB bandwidth is approximately 18 to 25 MHz. The intermodulation products can be filtered under large tone spacing. Therefore, when the two-tone spacing increases to 30 MHz, the IIP3 increases to 8 dBm and 5 dBm at frequencies of 7.6 GHz and 10.1 GHz, respectively. Later, the MSWF was biased with a constant magnetic field of 1460 Gauss which corresponds to a resonant frequency of 6.2 GHz. The out-of-band IIP3 was measured at 5.0, 7.6, and 10.3 GHz.

Since MSW are not able to be excited outside of the allowable frequency range, the MSWF are unable to create output intermodulation products via the YIG MSW waves. The measured IIP3 are 43 dBm, 41 dBm, and 44 dBm at 5.0, 7.6, and 10.3 GHz, respectively and are independent of tone spacing. This large IIP3 value highlights the resilience of this tunable filter to a strong out-of-band blocker. Out-of-band IIP was also measured with the bias magnetic field tuned to zero gauss. The IIP3 value was still in the range of 41 to 45 dBm at frequencies of 5.0, 7.6, and 10.3 GHz. This implies that the MSSW is not the primary source of out-of-band intermodulation products. Future studies with an improved IIP3 test setup or improved linearity in the aluminum transducers could achieve an increase in the out-of-band IIP3 of the tunable filters.

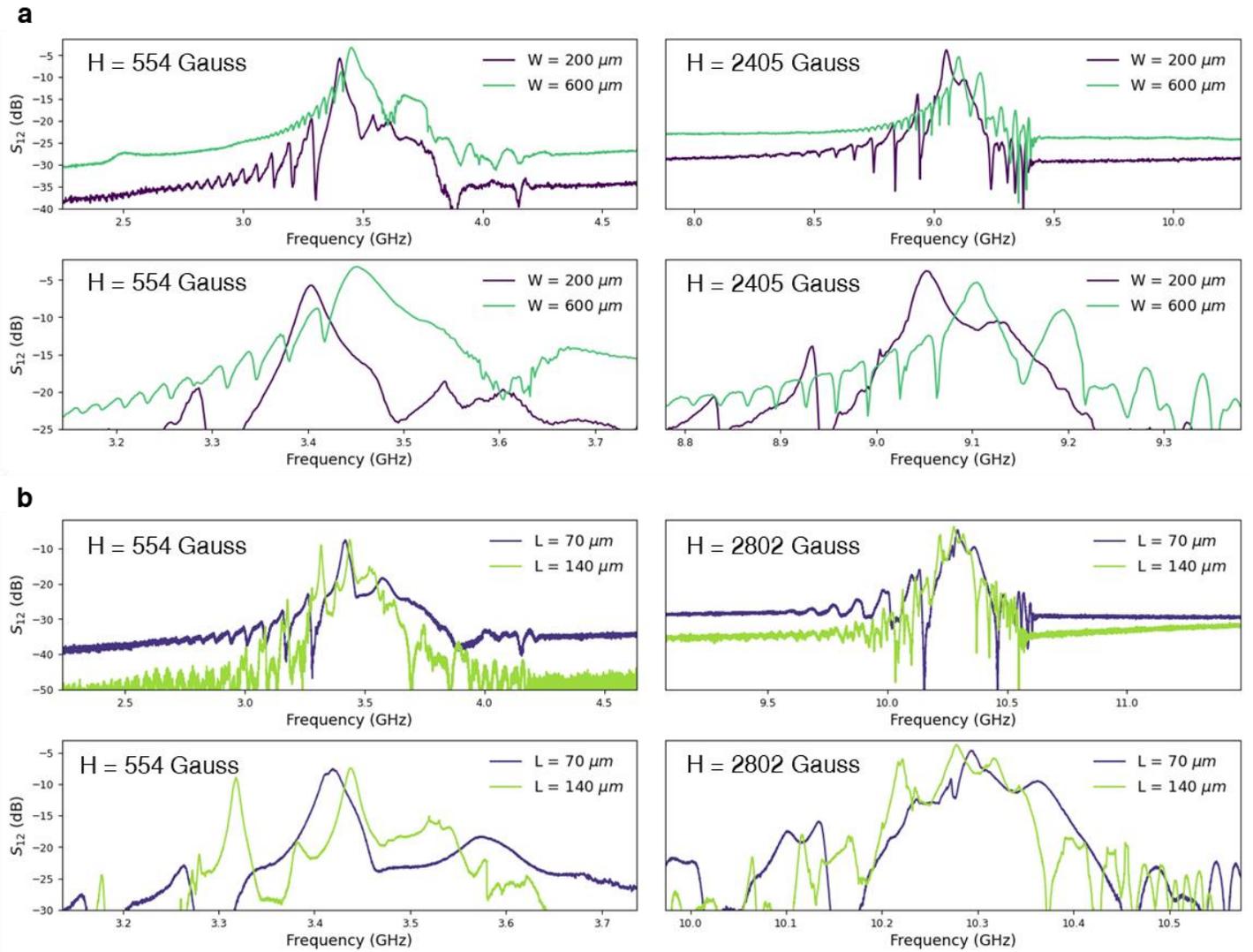

**Fig. 4 $S_{12}$ Frequency response of the MSWF at different magnetic flux density (H). a,** the impact of YIG cavity width on the frequency response with a constant length of 70 μm. **b,** the influence of YIG cavity length on the frequency response with a constant width of 150 μm.

## Magnetic Biasing Circuit

As shown in Fig. 1, the magnetic bias circuit comprises two neodymium-iron-boron (NdFeB) permanent magnets, two AlNiCo magnets wrapped with copper coils and two nickel-iron-molybdenum (NiFeMo) magnetic yokes. The NiFeMo magnetic yokes provide a low reluctance path for magnetic flux due to low coercivity and high permeability. The NdFeB permanent magnets and the coil-wound AlNiCo magnets serve as a constant magnetic flux source and a tunable magnetic flux source, respectively. Compared with NdFeB material, AlNiCo material has a lower coercivity[49]. Therefore, the AlNiCo magnets can be magnetized and demagnetized by applying a pulse of current though the coils surrounding the AlNiCo material. Also, due to the high magnetic remanence of AlNiCo, the AlNiCo magnets can still retain magnetism and provide magnetic flux for the circuit after the end of the current pulse.

To study the tuning range of the magnetic field, the magnetic flux density in the middle of the two yokes where the YIG chip sits was measured using a Gaussmeter. Supplementary Note 13 describes the process of capacitor charging and current flow through the coil to magnetize the AlNiCo magnets. By charging the capacitor to a specific voltage and discharging it through the coil, a current pulse is generated for magnetization. The remanent flux of the AlNiCo magnets can be adjusted by controlling the capacitor charging voltage and the resulting current pulse amplitude, allowing for precise tuning of the magnetic field. Fig. 5a shows the measured magnetic flux density along the vertical direction at different charging voltages.

The origin of the vertical position is where the magnetic flux density reaches the maximum value, which is around the middle of the yoke. Due to the 2 mm thickness of the magnetic yoke, the magnetic field does not change from -1 mm to 1 mm. This establishes a uniform magnetic bias field for the YIG filter cavity which is required for proper filter operation. Supplementary Note 14 describes a separate device in which a 0.5 mm thick yoke was utilized, resulting in a non-uniform magnetic field. This non-uniformity introduces additional losses in the integrated filter. With increasing capacitor charging voltage, the current for magnetizing the AlNiCo magnets increased, and thus the magnetic field generated by the AlNiCo magnets increased. The total amount of the magnetic flux in the middle of the two yokes was equal to the sum of the magnetic flux generated by the two NdFeB permanent magnets and the magnetic flux generated by the two AlNiCo magnets. For positive capacitor charging voltage, the field direction of the magnetized AlNiCo magnets was in the same direction as the NdFeB magnets. Due to the saturation of AlNiCo magnets, when the charging voltage was larger than 50 V, the effect of increasing the capacitor charging voltage on increasing the magnetic field became less. Finally, the tunable filter was realized by placing the MSWF chip in the center of the magnetic biasing circuit. About 0.7 J of energy is needed to switch from the minimum to the maximum bias field. The tunable filter assembly has a total dimension of 20 mm × 12 mm × 7 mm and occupies a volume of only 1.68 $cm^3$.

**Integrated device**

A MSWF with W = 200 μm and L = 70 μm with Al transducer width of 4 μm was laser diced and placed in the center of the gap of the magnetic biasing circuit. The filter was first measured inside a magnetic probe station with electromagnetic coils to provide the bias magnetic field. The frequency response was compared before and after integration with the tunable magnetic bias circuit. As shown in Fig. 5b-d, the $S_{12}$ frequency response and insertion loss remains unchanged between 3 GHz and 12 GHz in both the magnetic probe station and magnetic biasing circuit assemblies. The out-of-band rejection is greater than 25 dB and the insertion loss is less than 5.1 dB with an average across the tunable frequency range of 4 dB. The magnetic bias circuit measurements show a 7.7 GHz filter tuning range with a frequency tuning ratio of 3.3 achievable with an 80 V programming range. Fig. 5e presents detailed $S_{12}$ frequency responses zoomed around the passband of the MSWF. The filter exhibits a clean response without significant spurious or unwanted frequency components that could interfere with the desired signal.

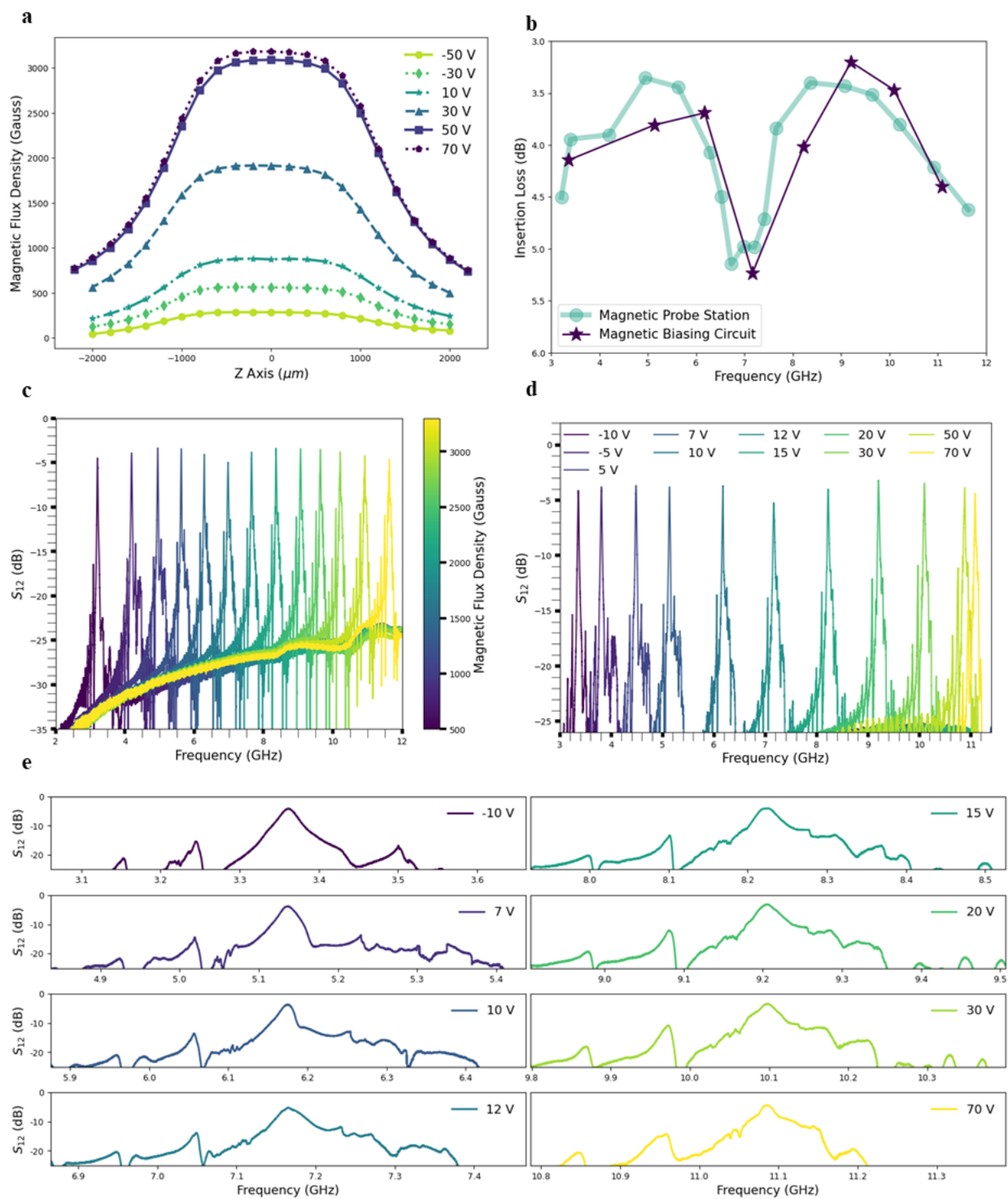

**Fig. 5 Integrated Device. a,** measured magnetic flux density under different capacitor charging voltages. **b,** insertion loss vs. frequency comparison for the MSWF measured under an external magnetic field generated by an electromagnet and the magnetic biasing circuit. **c,** MSWF response measured with the magnetic field

supplied by an electromagnet on a magnetic probe station. **d,** MSWF response measured with the magnetic field supplied by the zero DC power tunable magnetic biasing circuit. **e,** zoomed in $S_{12}$ frequency response of MSWF measured with magnetic field supplied by the tunable magnetic biasing circuit.

**Discussion**

In conclusion, we have demonstrated miniature and narrowband tunable filters with zero static power consumption, exceptional out-of-band linearity, and a frequency tuning range from 3.4 to 11.1 GHz. We revealed the tradeoff in width and length of the transducers in the design of MSSW cavity resonators. Additionally, we introduced a novel bias circuit utilizing NdFeB permanent magnets, coil-wound AlNiCo magnets, and NiFeMo magnetic yokes, which allows for magnetic field tuning through applied voltage pulses. This circuit exhibits promising potential for a diverse array of applications, offering electrically controlled magnetic fields with zero static power consumption.

A filter with a YIG cavity of W= 200 μm and L = 70 μm showed the high FoM, lowest insertion loss, and largest rejection of higher order modes and out-of-band signals. This new tunable filter technology exhibits immense promise across diverse domains, notably encompassing 5G and 6G cellular networks. In the realm of broadband Analog-to-digital converter (ADC) technology, the tunable filter's remarkable adaptability and ability to optimize the input spectrum play a crucial role. By addressing challenges posed by wideband ADCs, such as smaller available input voltage swing and therefore reduced dynamic range, our tunable filter ensures that wideband digital receivers stay within their dynamic range and handle data efficiently, even amid changing conditions. Moreover, for broadband antennas operating at frequencies from 3 to 11 GHz, our tunable filter's compact dimensions and wide frequency tuning range represent significant breakthroughs. Integrated wideband filtering simplifies antenna designs, allowing operation across a large bandwidth while leveraging the tunable filter for selective filtering. This approach facilitates efficient coexistence of various services on the ground while reducing the need for numerous antennas and line connections.[50]

The tunable filter's significance in 5G and 6G networks includes vital interference mitigation. For example, the sub-6 GHz 5G spectrum can overlap with C-band VSATs for Maritime and Fixed Satellite Services. This creates unpredictable 5G interferers, affecting users with adjacent-channel interference and Low-Noise Block (LNB) saturation.[50] Using our tunable filter in front of the LNB input can protect satellite carriers within the passband and isolate unwanted carriers, ensuring smooth operations and minimizing disruptions. The large reported out-of-band input third-order input intercept point (IIP3) of > +40 dBm allows higher levels of interfering signals before distortion. Future studies on a YIG films with larger thickness can further improve the linearity of the tunable filter.[48] In addition to the band pass filter illustrated in the paper, our filter platform can readily accommodate frequency tunable notch filtering by directly connecting two ports with a transducer on top of the YIG cavity. As illustrated in Supplementary Note 15, the MSWF will present a high impedance at resonance, effectively blocking unwanted signals at specific frequencies. This flexible and versatile design makes our tunable filter an effective device for interference management in advanced wireless networks like 5G and 6G.

Overall, the tunable filter's adaptability, wide frequency tuning range, low insertion loss, and zero static power consumption position it as a critical technology, effectively addressing challenges in broadband ADCs, broadband antennas, and interference mitigation in 5G and 6G networks. Its applications open new avenues for more efficient and dynamic RF front ends, ensuring optimal performance and seamless communication in the ever-evolving landscape of modern wireless technologies.

## Methods

**MSSW filters fabrication:** The YIG was grown using liquid epitaxy on a GGG substrate with <111> orientation (prepared by MTI corporation). A 100 nm thick $SiO_2$ was deposited as a hard mask using atomic layer deposition (Cambridge Nanotech S200) followed by 400 nm thick $SiO_2$ using plasma enhanced chemical vapor deposition (Oxford PlasmaLab 100). After annealing the sample to 600 °C in a nitrogen atmosphere, the hard mask layer was then patterned using standard photolithography and dry etching (Oxford 80 Plus RIE). The mask pattern was then transferred into the YIG using wet etching. Phosphoric acid at 140 °C was used to etch the YIG film with an etch rate of approximately 200 nm/min. The etch selectivity of YIG to $SiO_2$ using phosphoric acid was approximately 10:1. After patterning the YIG layer, the remaining $SiO_2$ layer was stripped using hydrofluoric acid. To pattern RF electrodes, 2 μm thick Al was deposited using sputtering at 1000 W with a base pressure of 1e-7 Torr (Evatec Clusterline 200 II) at 150 °C. The Al layer is patterned using standard photolithography and wet etching in Aluminum Etch Type A (Transene company, lnc) at 40 °C.

**Magnetic circuit fabrication:** The three different components of the magnetic bias circuit were prepared separately and then assembled. To form the magnetic yokes, a 2 mm thick NiFeMo sheet was cut using a 532 nm green laser to the required shape as shown in Fig. 1. The main body and protrusion part of the yokes have a size of 20 mm × 3 mm and 2 mm × 1.1 mm, respectively. The NdFeB permanent magnets had a dimension of 3.175 mm × 3.175 mm × 3.17 5mm and were purchased from K&J magnetics. The AlNiCo magnets had a dimension of 12 mm × 3 mm × 2 mm and were cut from a bulk AlNiCo bar using an electric discharge machining (EDM) copper wire with a diameter of 0.2 mm was wound around each AlNiCo magnet manually to achieve a total number of turns of 50. After all the magnetic parts were prepared, the two NiFeMo yokes and two NdFeB magnets were assembled and fixed on an acrylic substrate using epoxy. Then the coil-wound AlNiCo magnets were placed on the yoke and fixed using epoxy.

**Measurement setup:** The YIG sample was characterized using a magnetic probe station (MicroXact's MPS-1D-5kOe). The magnetic field was generated by electromagnets inside the magnetic probe station. A Gaussmeter (Model GM2, AlphaLab Inc) was used to calibrate the magnetic probe station and the filter frequency responses were measured using a vector network analyzer (Keysight, P9374A).

## Data availability

The data that support the findings of this study are available from the corresponding author upon reasonable request.

## Acknowledgement

The authors would like to thank Dr. Tim Hancock and Dr. Dave Abe of the Defense Advanced Research Projects Agency (DARPA) and Dr. Michael Page of the Air Force Research Laboratory for their guidance and support of this work under the DARPA Wideband Adaptive RF Protection (WARP) program, contract FA8650-21-1-7010. The fabrication of devices was performed at the Singh Center for Nanotechnology, supported by the NSF National Nanotechnology Coordinated Infrastructure Program (No. NNCI-1542153).

## Author contributions

X.D., M.I., F.A., M.A., and R.O. came up with the device concepts and experimental implementations. X.D., M.I., Y.D., T.Z., A.G., S.Y., J.P., F.A., M.A, and R.O. designed the devices and fabrication process flow. X.D. and M.I. fabricated the magnetostatic filters. Y.D., T.Z., and J.P. fabricated magnetic bias circuit. X.D., M.I., and S.Y. performed the measurements. X.D. and R.O. analyzed all data and wrote the manuscript. All authors have given approval to the final version of the manuscript.

## References

1. Yang B, Yu Z, Lan J, Zhang R, Zhou J, Hong W. Digital Beamforming-Based Massive MIMO Transceiver for 5G Millimeter-Wave Communications. *IEEE Transactions on Microwave Theory and Techniques* **66**, 3403-3418 (2018).

2. Wong PW, Hunter I. Electronically Tunable Filters. *IEEE Microwave Magazine* **10**, 46-54 (2009).

3. Ban Y, Liu J. An integrated tunable wideband RF filter in SiP technology. In: *2020 International Conference on Microwave and Millimeter Wave Technology (ICMMT)*) (2020).

4. Slovin G, El-Hinnawy N, Masse C, Rose J, Howard D. Multi-Throw SPNT Circuits Using Phase-Change Material RF Switches for 5G and Millimeter Wave Applications. In: *2021 IEEE MTT-S International Microwave Symposium (IMS)*) (2021).

5. Olsson RH, Bunch K, Gordon C. Reconfigurable Electronics for Adaptive RF Systems. In: *2016 IEEE Compound Semiconductor Integrated Circuit Symposium (CSICS)*) (2016).

6. Parke J*, et al.* High-Performance SLCFETs for Switched Filter Applications. In: *2016 IEEE Compound Semiconductor Integrated Circuit Symposium (CSICS)*) (2016).

7. Vetury R*, et al.* A Manufacturable AlScN Periodically Polarized Piezoelectric Film Bulk Acoustic Wave Resonator (AlScN P3F BAW) Operating in Overtone Mode at X and Ku Band. In: *2023 IEEE/MTT-S International Microwave Symposium - IMS 2023*) (2023).

8. Izhar*, et al.* A K-Band Bulk Acoustic Wave Resonator Using Periodically Poled $Al_{0.72}Sc_{0.28}N$. *IEEE Electron Device Letters* **44**, 1196-1199 (2023).

9. Kramer J*, et al.* 57 GHz Acoustic Resonator with k2 of 7.3 % and Q of 56 in Thin-Film Lithium Niobate. In: *2022 International Electron Devices Meeting (IEDM)*) (2022).

10. Lu R, Yang Y, Link S, Gong S. Enabling Higher Order Lamb Wave Acoustic Devices With Complementarily Oriented Piezoelectric Thin Films. *Journal of Microelectromechanical Systems* **29**, 1332-1346 (2020).

11. Al-Yasir YI, Ojaroudi Parchin N, Abd-Alhameed RA, Abdulkhaleq AM, Noras JM. Recent progress in the design of 4G/5G reconfigurable filters. *Electronics* **8**, 114 (2019).

12. Psychogiou D, Gómez-García R, Peroulis D. Recent advances in reconfigurable microwave filter design. In: *2016 IEEE 17th Annual Wireless and Microwave Technology Conference (WAMICON)*). IEEE (2016).

13. Rebeiz GM*, et al.* Tuning in to RF MEMS. *IEEE Microwave Magazine* **10**, 55-72 (2009).

14. Mitola J, Maguire GQ. Cognitive radio: making software radios more personal. *IEEE Personal Communications* **6**, 13-18 (1999).

15. Ghadaksaz M. Novel active RF tracking notch filters for interference suppression in HF, VHF, and UHF frequency hopping receivers. In: *MILCOM 91 - Conference record*) (1991).

16. Laplanche E*, et al.* Tunable Filtering Devices in Satellite Payloads: A Review of Recent Advanced Fabrication Technologies and Designs of Tunable Cavity Filters and Multiplexers Using Mechanical Actuation. *IEEE Microwave Magazine* **21**, 69-83 (2020).

17. Fischer G. Next-generation base station radio frequency architecture. *Bell Labs Technical Journal* **12**, 3-18 (2007).

18. Tsui JB. *Microwave Receivers with Electronic Warfare Applications*. Institution of Engineering and Technology (2005).

19. Brown JA, Barth S, Smyth BP, Iyer AK. Compact Mechanically Tunable Microstrip Bandstop Filter With Constant Absolute Bandwidth Using an Embedded Metamaterial-Based EBG. *IEEE Transactions on Microwave Theory and Techniques* **68**, 4369-4380 (2020).

20. Kurudere S, Ertürk VB. Novel Microstrip Fed Mechanically Tunable Combline Cavity Filter. *IEEE Microwave and Wireless Components Letters* **23**, 578-580 (2013).

21. Basavarajappa G, Mansour RR. Design Methodology of a Tunable Waveguide Filter With a Constant Absolute Bandwidth Using a Single Tuning Element. *IEEE Transactions on Microwave Theory and Techniques* **66**, 5632-5639 (2018).

22. Sinanis MD, Adhikari P, Jones TR, Abdelfattah M, Peroulis D. High-Q High Power Tunable Filters Manufactured With Injection Molding Technology. *IEEE Access* **10**, 19643-19653 (2022).

23. Liu X, Katehi LPB, Chappell WJ, Peroulis D. High-Q Tunable Microwave Cavity Resonators and Filters Using SOI-Based RF MEMS Tuners. *Journal of Microelectromechanical Systems* **19**, 774-784 (2010).

24. Yang Z, Zhang R, Peroulis D. Design and Optimization of Bidirectional Tunable MEMS All-Silicon Evanescent-Mode Cavity Filter. *IEEE Transactions on Microwave Theory and Techniques* **68**, 2398-2408 (2020).

25. Yang T, Rebeiz GM. Bandpass-to-Bandstop Reconfigurable Tunable Filters with Frequency and Bandwidth Controls. *IEEE Transactions on Microwave Theory and Techniques* **65**, 2288-2297 (2017).

26. Huang F, Fouladi S, Mansour RR. High-Q Tunable Dielectric Resonator Filters Using MEMS Technology. *IEEE Transactions on Microwave Theory and Techniques* **59**, 3401-3409 (2011).

27. Courreges S, Li Y, Zhao Z, Choi K, Hunt A, Papapolymerou J. A Low Loss X-Band Quasi-Elliptic Ferroelectric Tunable Filter. *IEEE Microwave and Wireless Components Letters* **19**, 203-205 (2009).

28. Wei Z, Yang T, Chi PL, Zhang X, Xu R. A 10.23–15.7-GHz Varactor-Tuned Microstrip Bandpass Filter With Highly Flexible Reconfigurability. *IEEE Transactions on Microwave Theory and Techniques* **69**, 4499-4509 (2021).

29. Mohammadi L, Koh K-J. 2–4 GHz Q-tunable LC bandpass filter with 172-dBHz peak dynamic range, resilient to+ 15-dBm out-of-band blocker. In: *2015 IEEE Custom Integrated Circuits Conference (CICC)*). IEEE (2015).

30. Luo Ck, Buckwalter JF. A 0.25-to-2.25 GHz, 27 dBm IIP3, 16-Path Tunable Bandpass Filter. *IEEE Microwave and Wireless Components Letters* **24**, 866-868 (2014).

31. Dai S, Bhave SA, Wang R. Octave-Tunable Magnetostatic Wave YIG Resonators on a Chip. *IEEE Transactions on Ultrasonics, Ferroelectrics, and Frequency Control* **67**, 2454-2460 (2020).

32. Prabhakar A, Stancil DD. *Spin waves: Theory and applications*. Springer (2009).

33. Ishak WS. Magnetostatic wave technology: a review. *Proceedings of the IEEE* **76**, 171-187 (1988).

34. Mahmoud A, *et al.* Introduction to spin wave computing. *Journal of Applied Physics* **128**, 161101 (2020).

35. Marcelli R, Gasperis PD, Marescialli L. A tunable, high Q magnetostatic volume wave oscillator based on straight edge YIG resonators. *IEEE Transactions on Magnetics* **27**, 5477-5479 (1991).

36. Tsai CS, Qiu G. Wideband Microwave Filters Using Ferromagnetic Resonance Tuning in Flip-Chip YIG-GaAs Layer Structures. *IEEE Transactions on Magnetics* **45**, 656-660 (2009).

37. Röschmann P. YIG filters. *Philips Tech Rev* **32**, 322-327 (1971).

38. Aigle M, Hechtfischer G, Hohenester W, Junemann R, Evers C. A systematic way to YIG-filter-design. In: *2007 European Microwave Conference*). IEEE (2007).

39. Castéra JP, Hartemann P. Magnetostatic wave resonators and oscillators. *Circuits, Systems and Signal Processing* **4**, 181-200 (1985).

40. Zhu Y, Qiu G, Chi KH, Wang BBT, Tsai CS. A Tunable X-Band Band-Pass Filter Module Using YIG/GGG Layer on RT/Duroid Substrate. *IEEE Transactions on Magnetics* **45**, 4195-4198 (2009).

41. Du S, Yang Q, Wang M, Fan X. A Magnetically Tunable Bandpass Filter With High Out-of-Band Suppression. *IEEE Transactions on Magnetics* **58**, 1-5 (2022).

42. Yang GM, Wu J, Lou J, Liu M, Sun NX. Low-Loss Magnetically Tunable Bandpass Filters With YIG Films. *IEEE Transactions on Magnetics* **49**, 5063-5068 (2013).

43. Chen J, Yu H, Gubbiotti G. Unidirectional spin-wave propagation and devices. *Journal of Physics D: Applied Physics* **55**, 123001 (2021).

44. Wong KL, *et al.* Unidirectional propagation of magnetostatic surface spin waves at a magnetic film surface. *Applied Physics Letters* **105**, 232403 (2014).

45. Ganguly AK, Webb DC. Microstrip Excitation of Magnetostatic Surface Waves: Theory and Experiment. *IEEE Transactions on Microwave Theory and Techniques* **23**, 998-1006 (1975).

46. Vanderveken F, Tyberkevych V, Talmelli G, Sorée B, Ciubotaru F, Adelmann C. Lumped circuit model for inductive antenna spin-wave transducers. *Scientific Reports* **12**, 3796 (2022).

47. O'Keeffe TW, Patterson RW. Magnetostatic surface-wave propagation in finite samples. *Journal of Applied Physics* **49**, 4886-4895 (1978).

48. Adam JD, Winter F. Magnetostatic wave frequency selective limiters. *IEEE transactions on magnetics* **49**, 956-962 (2013).

49. Jiles D. *Introduction to magnetism and magnetic materials*. CRC press (2015).

50. De Paolis F. Satellite filters for 5G/6G and beyond. In: *2021 IEEE MTT-S International Microwave Filter Workshop (IMFW)*). IEEE (2021).

# Frequency Tunable Magnetostatic Wave Filters With Zero Power Magnetic Biasing Circuitry
# Supplementary Information

Xingyu Du[1], Mohamad Hossein Idjadi[1], Yixiao Ding[1], Tao Zhang[1], Alexander J. Geers[1], Shun Yao[1], Jun Beom Pyo[1], Firooz Aflatouni[1], Mark Allen[1], and Roy H. Olsson III[1*]

[1]Department of Electrical and Systems Engineering, University of Pennsylvania, Philadelphia, PA, USA
Email: rolsson@seas.upenn.edu

# Table of Contents

**Supplementary Note 1: Comparison of performance metrics for electrically continuous tunable bandpass filters.**

| Type | Freq. Range (GHz) | Tuning Ratio | In-Band Insertion Loss (dB) | Out-of-band Rejection (dB) | Out-of-Band 3rd order input referred inter-modulation intercept point (OOB IIP3) (dBm) | Power Consumption | Size (cm$^3$) |
|---|---|---|---|---|---|---|---|
| This work (Planar YIG) | 3.4-11.1 | 3.3:1 | 3.2-5.1 | >25 | >41 | 0 | 1.68 |
| Planar YIG[1] | 2-12 | 6:1 | 20-32 | >40 | Not reported | Electro-magnet (>58 W[2])* | Electro-magnet (75-99 [2])* |
| Planar YIG[3] | 5.2-7.5 | 1.4:1 | 1.6-3 | >20 | Not reported | Electro-magnet (>58 W[2])* | Electro-magnet (75-99 [2])* |
| Planar YIG[4] | 11-16 | 1.5:1 | ~3.5 | ~60 | Not reported | Electro-magnet (>58 W[2])* | Electro-magnet (75-99[2])* |
| YIG Sphere[5] | 2-18 | 9:1 | <6 | >80 | Not reported | Electro-magnet (2-10 W) | Electro-magnet (22.7) |
| Varactor[6] | 1.6-2.1 | 1.3:1 | 2.4-3 | >50 | Not reported | Microwatts[7] | 0.2 |
| Varactor[8] | 10.2-15.7 | 1.5:1 | 2.6-3.9 | >35 | Not reported | Microwatts[9] | 0.4 |
| Varactor[10] | 8.4-9.2 | 1.1:1 | 3.5-5.7 | ~30 | Not reported | Microwatts[11] | 0.05 |
| Varactor[12] | 2.4-2.9 | 1.2:1 | 3.3-5.1 | >10 | Not reported | 12 µW | 0.05 |
| RF MEMS[13] | 2.8-5.2 | 1.8:1 | 0.1 | >40 | >60 | 0 | ~80 |
| RF MEMS[14] | 3.0-4.7 | 1.6:1 | 2-4 | >50 | Not reported | 0 | ~8 |

Supplementary Table 1. Performance metrics comparison for microwave tunable bandpass filters spanning from S band to X band (2-12 GHz). The indicated size encompasses the complete assembly of the filter, incorporating electromagnets for YIG-based filters.

\*: These papers reference the use of electromagnets to generate the external magnetic field required for the YIG material, yet they do not elaborate on the specific design of these electromagnets. However, in a related study conducted by Chen S. Tsai and Gang Qiu, details regarding the electromagnet dimensions and specifications are provided. The total electromagnet dimensions were approximately 63.5 mm (height) by 19.1 mm (width) by 61.8-81.4 mm (length), resulting in a volume range of 75-99 cm$^3$. The magnetic circuit employed two solenoids, each requiring a 1.2 A current and possessing a resistance of 20 Ω to generate a magnetic field of about 300 Gauss. Consequently, the power consumption for this electromagnet setup exceeded 58 W.

In Supplementary Figure 1, a size comparison between this work and commercial YIG spheres-based tunable filters, namely Teledyne F series and Micro Lambda Wireless MLFR series, is illustrated.[5, 15] The images are depicted at the same scale. The YIG sphere-based filters are notably larger due to the electromagnets required to adjust their resonance frequency.

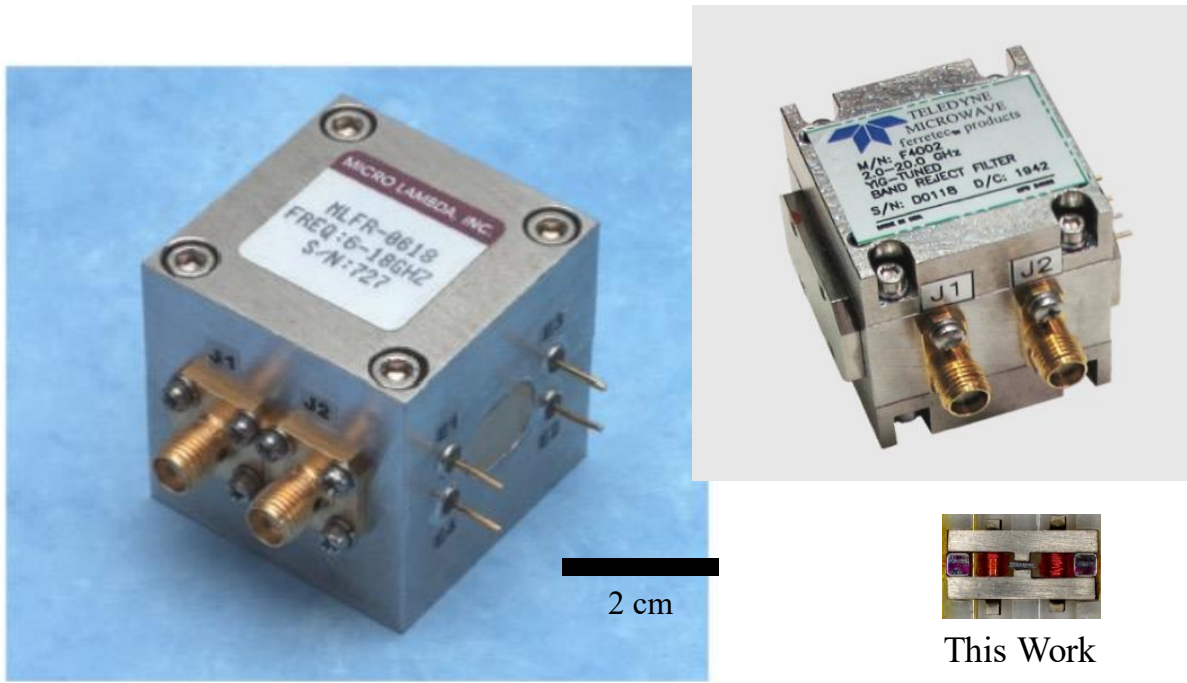

Supplementary Figure 1. A size comparison between this work and commercial YIG sphere-based tunable filters, namely Teledyne F4002 and Micro Lambda Wireless MLFR-0618.

**Supplementary Note 2: Overall Layout of the Magnetostatic Wave Resonators (MSWR)**

Supplementary Figure 2 depicts the arrangement of the MSSW devices. The measurement employed two pairs of Ground-Signal-Ground (GSG) probes with a pitch of 150 μm (GGB, 40A-GSG-150-P model). The ground pads possess a width of 130 μm, while the source pads have a width of 110 μm. To establish connectivity between the two ground pads, an Aluminum (Al) trace with a width of 60 μm is utilized. Beyond the signal contact pads, the Al traces undergo a reduction in width. This reduction achieves a minimum width of 5 μm in the Aluminum transducer nearest to the YIG regions, which are the coupling regions. The coupling efficiency with the magnetostatic wave is directly proportional to the current density. To address this, taper structures are incorporated to convert the 5 μm width of the coupling Al transducer to wider contact pads.

COMSOL simulations have been utilized to simulate the structure. In the absence of an external magnetic field, an isolation ($S_{12}$) of more than 25 dB can be achieved. The series resistance and inductance have been optimized by employing an aluminum thickness of 2 μm. The series components and isolation characteristics are further analyzed in the subsequent sections.

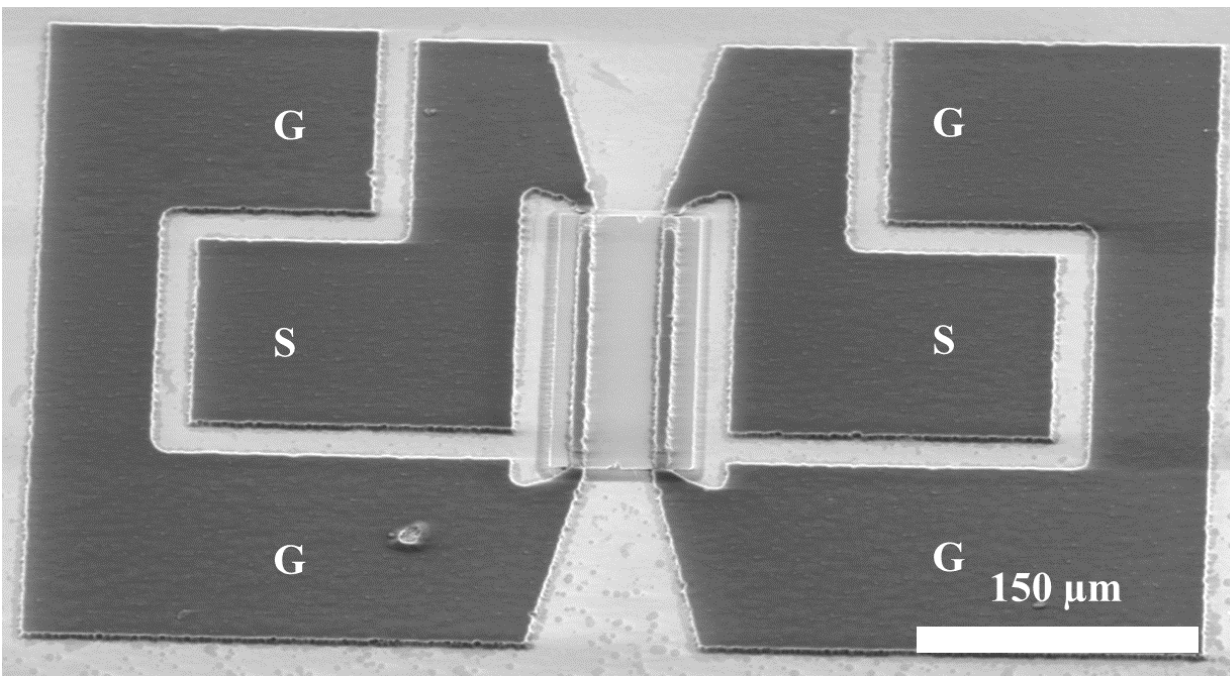

Supplementary Figure 2. SEM image of MSWF with the transducers on top of the YIG setup with width of 150 μm and length of 70 μm. Ground (G) and Signal (S) contact pads are shown in the image.

**Supplementary Note 3: Magnetic Probe Station Measurement Setup**

Supplementary Figure 3 illustrates the experimental configuration utilized for measuring the MSSW filters. An array of MSSW filters were fabricated on a YIG/GGG (Yttrium Iron Garnet/Gadolinium Gallium Garnet) substrate and positioned on a 5 mm thick acrylic spacer. The Acrylic spacer was centrally placed on a metallic stage. To ensure a consistent and adjustable magnetic field across the MSSW filters, two electromagnets were positioned on either side of the metallic stage. The magnitude of the magnetic field was regulated by controlling the current within the coils of the electromagnets, employing closed loop sensor feedback control for precise adjustments.

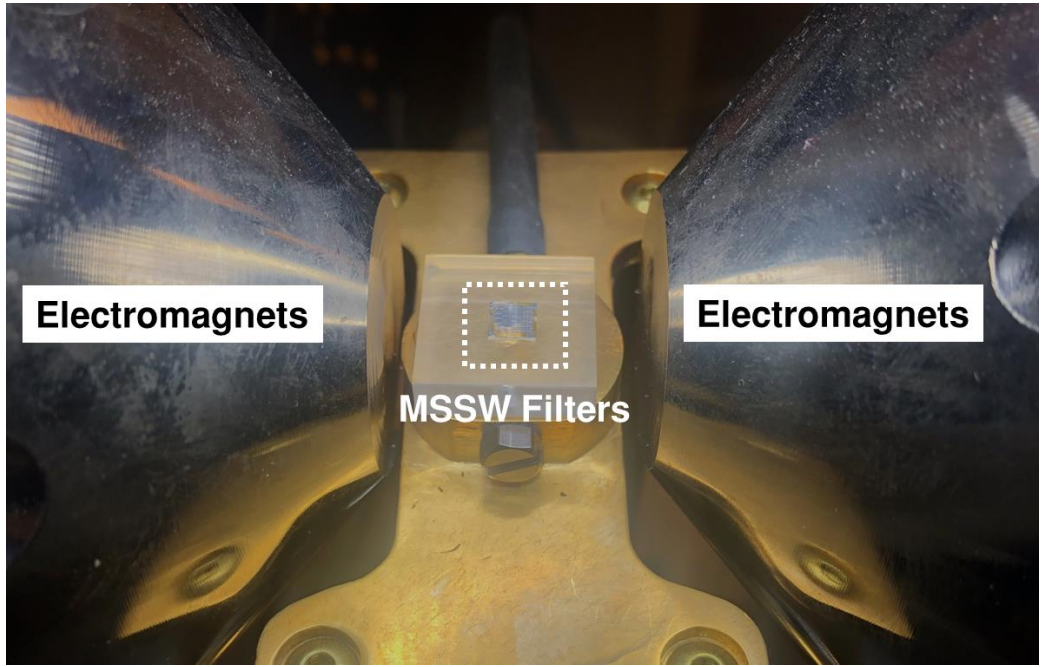

Supplementary Figure 3. MSSW filters inside the magnetic probe station measurement setup.

**Supplementary Note 4: Circuit Modeling for Magnetostatic Wave Resonators (MSWR)**

Following the fabrication and measurement of the two-port MSWF, using the magnetic probe station, the performance of the fabricated MSWR and MSWF were evaluated by studying the circuit model of the devices. In the measurement of the MSWF, the full S-parameters of the two ports were recorded. To determine the characteristics of the one-port MSWR device, we utilized the values obtained from either $Z_{11}$ or $Z_{22}$ of the MSWR, representing the impedance when the other port is open.

As outlined in Supplementary Note 6, the MSSW filters exhibit multiple modes owing to the highly dispersive nature of spin waves in a planar rectangular YIG cavity. Consequently, these different modes are closely clustered together in frequency. Unlike spurious-free single mode resonators, the frequency response of MSSW filters can display numerous local maxima and minima. The anticipated behavior of an ideal resonance, characterized by maximum impedance at the resonance frequency ($f_s$) and minimum at anti-resonance frequency ($f_p$), becomes challenging to observe. Consequently, the accurate extraction of circuit parameters for MSWR becomes a laborious task. This issue is less pronounced in devices with low coupling ($K^2$), primarily because the narrow frequency spacing between $f_s$ and $f_p$ results in fewer spurious modes within that range.

In a conventional acoustic resonator, a multi-mode Modified Butterworth-Van Dyke (MBVD) circuit can be established by introducing several parallel branches of the motional resistor $R_x$, the motional inductor $L_x$, and the motional capacitor $C_x$ to the main mode resonance tank. The circuit parameters derived from these additional parallel branches enable the calculation of the coupling coefficient and quality factor of the spurious mode.[16] Inspired by the multi-mode MBVD circuit, Figure 2 illustrates the circuit diagram of MSWR with a total of p modes. In each resonance tank (designated as the nth tank in the circuit), there is a parallel magnetostatic resistor $R_m$, one magnetostatic inductance $L_m$, and one magnetostatic capacitor $C_m$. These resonance tanks are connected in series, the impedance of the resonance tank in the MSWR reaches its maximum at resonance and returns to a minimum at frequencies away from the resonance. This behavior contrasts with traditional acoustic resonators, where additional branches would remain open when only the main branch is resonating. In MSSW resonators, the additional branches remain electrically shorted when only the main branch is resonating. This unique property of MSWR leads to the $S_{11}$ parameter of both the MSWR and MSWF exhibiting a bandstop characteristic, indicating the $Z_{11}$ impedance reaches the maximum at resonance frequency. Additionally, the $S_{12}$ parameter of the MSWF shows a bandpass characteristic, indicating the $Z_{12}$ impedance reaches the maximum at resonance frequency.

As for a single mode MSWR, the impedance of a resonator is the series combination of series resistor $R_s$ and series inductance $L_s$ and the resonance tank. It can be expressed as follows.

$$Z_{single} = Z_{series} + Z_{tank\_single} = R_s + j\omega L_s + \frac{1}{\frac{1}{R_m} + \frac{1}{j\omega L_m} + j\omega C_m} \quad \text{(S1)}$$

The total impedance of an MSWR's tank consisting of a total of $p$ modes can be expressed by combining the impedance of each individual resonance tank, denoted as: $Z_{tank1}$, $Z_{tank2}$, $Z_{tank3}$, ......, $Z_{tankp}$. Consequently, the total impedance of a MSWR, $Z_{total}$, can be described as follows:

$$Z_{total} = Z_{series} + \sum_{n=1}^{p} Z_{tank\_n} = R_s + j\omega L_s + \sum_{n=1}^{p} \frac{1}{\frac{1}{R_{m1n}} + \frac{1}{j\omega L_{m1n}} + j\omega C_{m1n}} \tag{S2}$$

With the establishment of the multi-mode circuit model for MSW devices, the process flow employed to extract the circuit parameters is described in Supplementary Fig. 4. The determination of these parameters assumes a critical role in precisely characterizing the device's behavior and facilitating performance optimization.

The procedure commences by importing the measured impedance data, $Z_{mea}$. Subsequently, the $R_s$ and $L_s$ are calculated from the imaginary and real parts of the off-resonance impedance, respectively. When the frequency significantly deviates from the resonance, the overall $Z_{11}$ impedance of MSWR can be expressed as, $Z_{series}$.

$$Z_{series} = R_{s1} + j\omega L_{s1} \tag{S3}$$

Subsequently, the impedance response is analyzed to identify the frequency peaks of interest. These peaks, usually corresponding to local maxima with the highest magnitude, are then ranked based on their impedance values. For instance, the first peak, representing the resonance with the largest impedance, is used to calculate the circuit parameters of the resonance tank associated with the largest. The second peak represents the second-largest resonance impedance, and so on. By examining the resonance peaks in descending order, it becomes possible to observe and analyze the impedance changes of smaller resonance tanks, as the significant impedance of larger resonance modes can conceal the impedance response of smaller modes. To accurately determine the fitting frequency range, a local minimum preceding the selected peak and a local minimum following it are identified. This frequency range is crucial as it captures the dominant influence of the corresponding resonance tank's impedance on the overall circuit behavior. By focusing on this range, a precise characterization of the specific resonance mode can be achieved.

After that, the initial guess parameter for the resonance tank circuit parameters: $R_m$, $L_m$, and $C_m$ can been estimated for the initialization of the multi-mode circuit model recursive fitting. The initial condition for $R_m$ is the frequency response's maximum magnitude of the impedance. The Q can be approximated as the ratio of $f_s$ over this peak's 3-dB bandwidth. The initial condition for $L_m$ and $C_m$ can be expressed as:

$$L_m = \frac{R_m}{2\pi f_s Q} \tag{S4}$$

$$C_m = \frac{1}{L_m (2\pi f_s)^2} \tag{S5}$$

After obtaining the initial values for $R_m$, $L_m$, and $C_m$, the random search process is employed to explore the parameter space for these three circuit parameters. This process begins by generating a series of three sets of random numbers that deviate from the initial three parameters. These random numbers serve as the test parameters for the subsequent analysis. By incorporating these test parameters into the circuit model, the corresponding impedance response of $Z_{fit}$ can be computed as the combination of the impedance of the series components with the impedance of the previously fit resonance tank, if any, and the resonance tank of interest with random generated test parameters.

In order to compare between the calculated impedance response and the measured impedance data, an error function of $\sigma(Z_{fit}, Z_{mea})$ can be calculated from the magnitude of the fit data, $dB(Z_{fit})$ and the measured data, $dB(Z_{mea})$ and the phase of the fit data, $phase(Z_{fit})$ and the measured data, $phase(Z_{mea})$, which is defined as:

$$\sigma(Z_{fit}, Z_{mea}) = \sigma_{dB} + \sigma_{phase} \\ = |Norm(dB(Z_{fit})) - Norm(dB(Z_{mea}))| \\ + |Norm(phase(Z_{fit})) - Norm(phase(Z_{mea}))| \tag{S6}$$

The normalization function, denoted as $Norm(x)$, serves to eliminate the absolute scale difference between phase and magnitude. It can be mathematically expressed as follows:

$$Norm(x) = \frac{x - mean(x)}{std(x)} \tag{S7}$$

In this equation, $x$ is the data of interest. The $mean(x)$ and $std(x)$ are the average and standard deviation of the data.

The error function is utilized to evaluate all the test parameters generated during the random search process. The test parameters associated with the smallest error function value are saved, as they indicate a closer match between the modeled impedance response and the measured data. This process of random number generation and error function evaluation is repeated multiple times until the error function converges, indicating a satisfactory level of parameter optimization. Once the error convergence is achieved, the circuit extraction process concludes with key fitting results for $R_m$, $L_m$, and $C_m$. These results represent the optimized circuit parameters that best align with the measured impedance data. Subsequently, the recursive fitting procedure moves on to the next peak of interest and performs the fitting for the resonance tank associated with the next mode. This iterative approach ensures comprehensive characterization of each resonance mode and the determination of their respective circuit parameters.

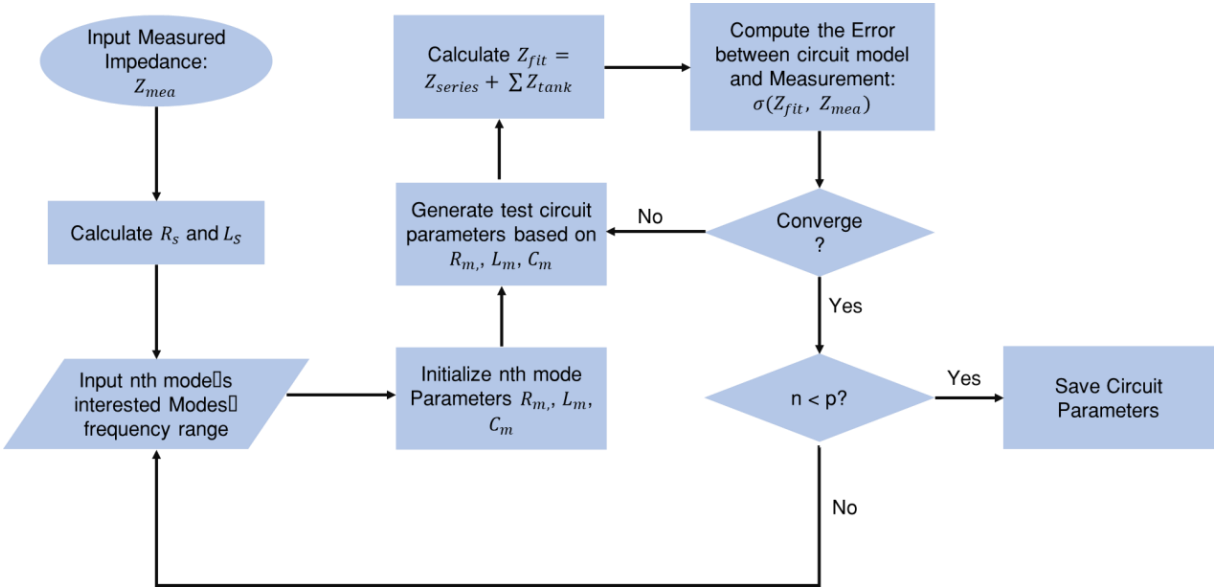

Supplementary Figure 4. Multi-resonance recursive fitting scheme for accurate extraction of the MSWR model with multiple resonance models.

In Figure 2, a comparison of the multi-mode circuit model, single-mode circuit model, and measurement are presented. Supplementary Figure 5 illustrates the details of each resonance tank in the multi-mode circuit model. In this MSWR with width of 200 μm and length of 70 μm, the top five local maximum are selected and the multi-mode recursive fitting was performed with five resonance tanks. By systematically generating and evaluating test parameters, this approach successfully identifies the parameter values that result in the closest agreement between circuit model and the measured data. The series $R_s$ is 0.88 Ω and the series $L_s$ is 0.3436 nH. The circuit parameters for the resonance tanks are shown in Supplementary Table 2.

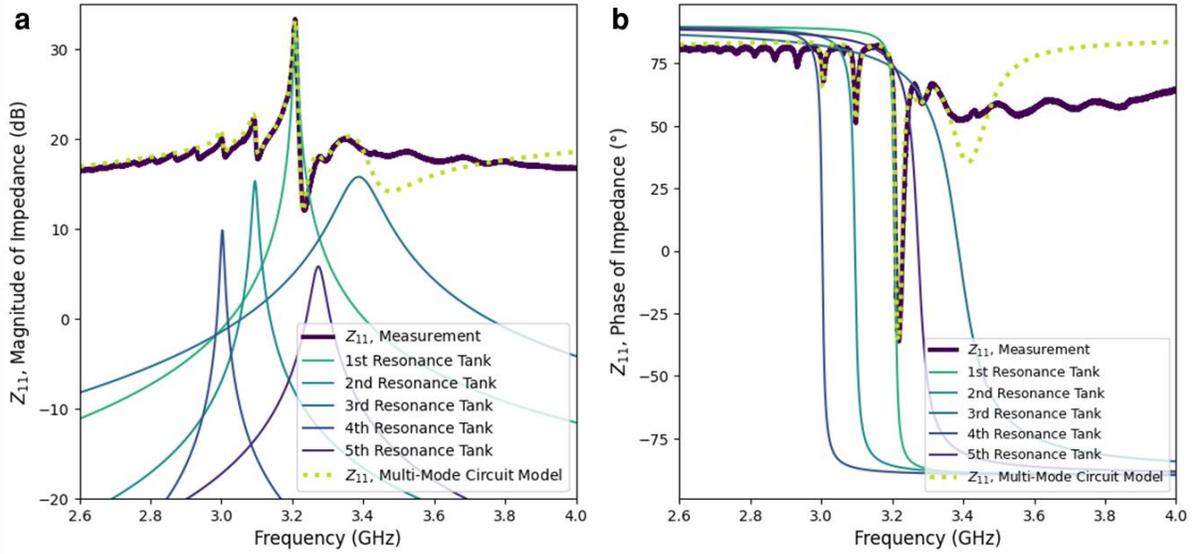

Supplementary Figure 5. Example of the multi-resonance recursive fitting using measurement of typical MSWR with width of 200 µm and length of 70 µm with transducer width of 4 µm. (a) Magnitude comparison between the measurement and the circuit model. (b) Phase comparison between the measurement and the circuit model.

|  | Resonance Tank #1 | Resonance Tank #2 | Resonance Tank #3 | Resonance Tank #4 | Resonance Tank #5 |
|---|---|---|---|---|---|
| Resonance Frequency (GHz) | 3.2 | 3.1 | 3.3 | 3.0 | 3.3 |
| $R_m$ (Ω) | 42 | 5.8 | 6.1 | 3.1 | 1.9 |
| $L_m$ (pH) | 5.8 | 1.4 | 9.7 | 0.56 | 1.2 |
| $C_m$ (nF) | 0.4 | 1.9 | 0.23 | 5.0 | 2.0 |

Supplementary Table 2. Summary of the circuit parameters for multi-mode circuit model of MSWR with W = 200 µm, L = 70 µm, and transducer width of 4 µm.

Another application for the circuit model is to understand the effect of series resistance and inductance, as the spin wave Q-factor can be calculated from the circuit model. The spin wave Q-factor is defined as:

$$Spin\ Wave\ Q = \frac{R_m}{2\pi L_m} \quad (S8)$$

Supplementary Figure 6 illustrates the spin wave Q-factor of various MSWR with different widths. The spin wave Q-factor and device Q-factor follow similar trends with frequency. The maximum spin wave Q-factor of 1777 for W = 150 μm, L = 70 μm at 11.5 GHz can be achieved. Compared to device Q-factor, spin wave Q-factor removes the impact of the Al trace resistance on overall device performance and is about 30 % ~ 50% higher than the device Q. This indicates significant energy is dissipated through the series resistance.

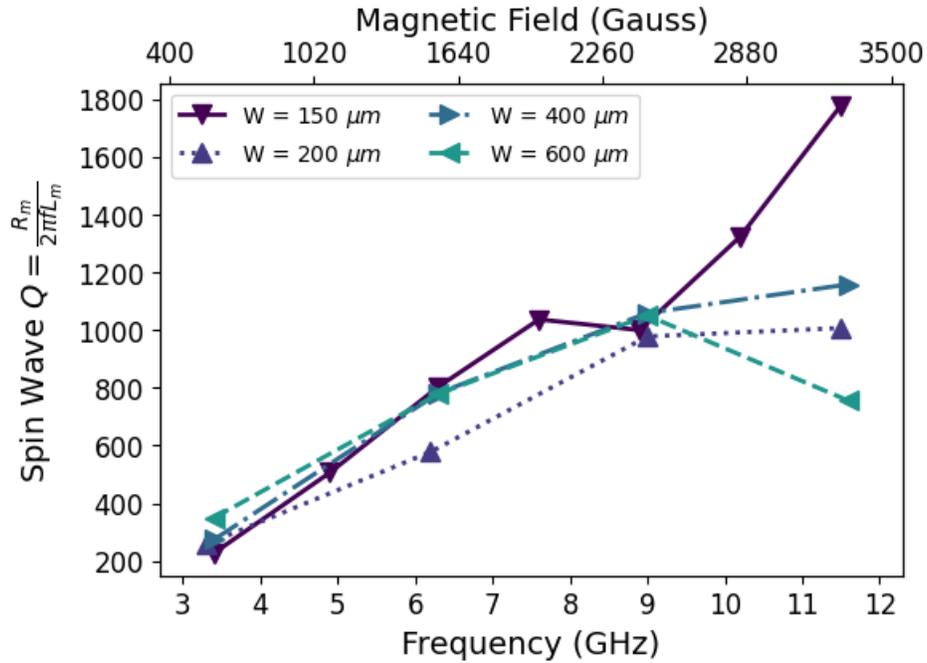

Supplementary Figure 6. Comparison of MSWR spin wave Q-factor for different width of YIG cavity.

**Supplementary Note 5: Width Effect of the Aluminum Transducers**

In order to achieve a low insertion over a broad frequency tuning range, the effect of the width of the aluminum (Al) transducers has been studied. A 2-dimensional finite element simulation was performed in COMSOL [17] to simulate the total integration of the magnetic flux density in the YIG when a DC current is applied to the aluminum trace line. The aluminum transducer was simulated with a trapezoidal shape to better represent the wet etching profile. The change in series resistance of the aluminum transducers with Al width is much less than 1 Ω when the width of the aluminum transducer decreases from 19 μm to 4 μm. Thus, a constant current assumption of 1 A was used for the simulation. Supplementary Figure 7 shows that a narrower Al transducer achieves higher magnetic flux, which is mainly due to the increase of the y component of the magnetic flux.

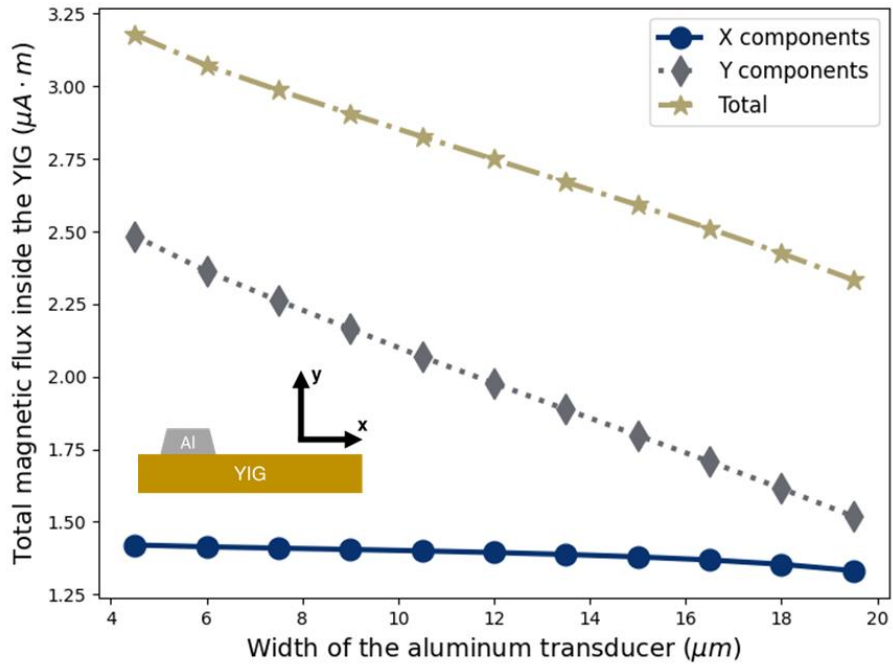

Supplementary Figure 7. COMSOL simulation results on the magnetic flux inside the YIG with respect to the width of the aluminum transducer.

As a result of the increased magnetic field inside the YIG cavity, the $R_m$ and insertion significantly improve between 3.4 ~ 12.9 GHz, as shown in Supplementary Figure 8. The increased FOM does not translate into an improvement in insertion loss for frequencies above 12.9 GHz because the MSWR resonator is over coupled to the 50 Ω source impedance.

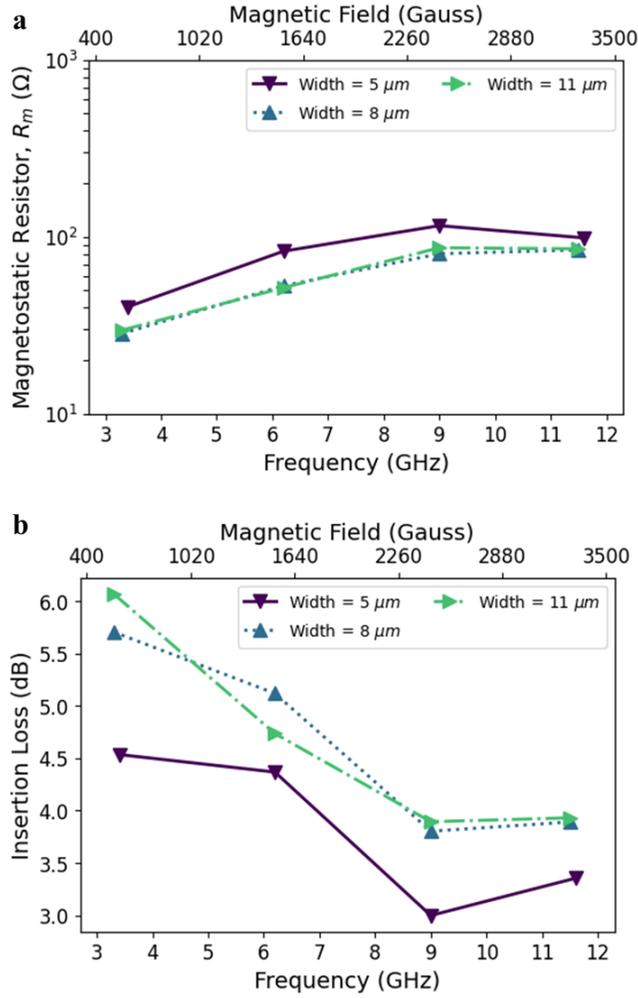

Supplementary Figure 8. Effect of the aluminum transducer width on the (a) radiation impedance and (b) insertion loss. The MSWF is designed with constant W = 200 μm and L = 70 μm.

Supplementary Figure 9 shows a typical frequency response for different Al transducer widths. A narrower aluminum transducer slightly reduces the out-of-band attenuation of the tunable filters as the increase of the current density in the aluminum trace slightly enhances the direct inductive coupling between the two ports.

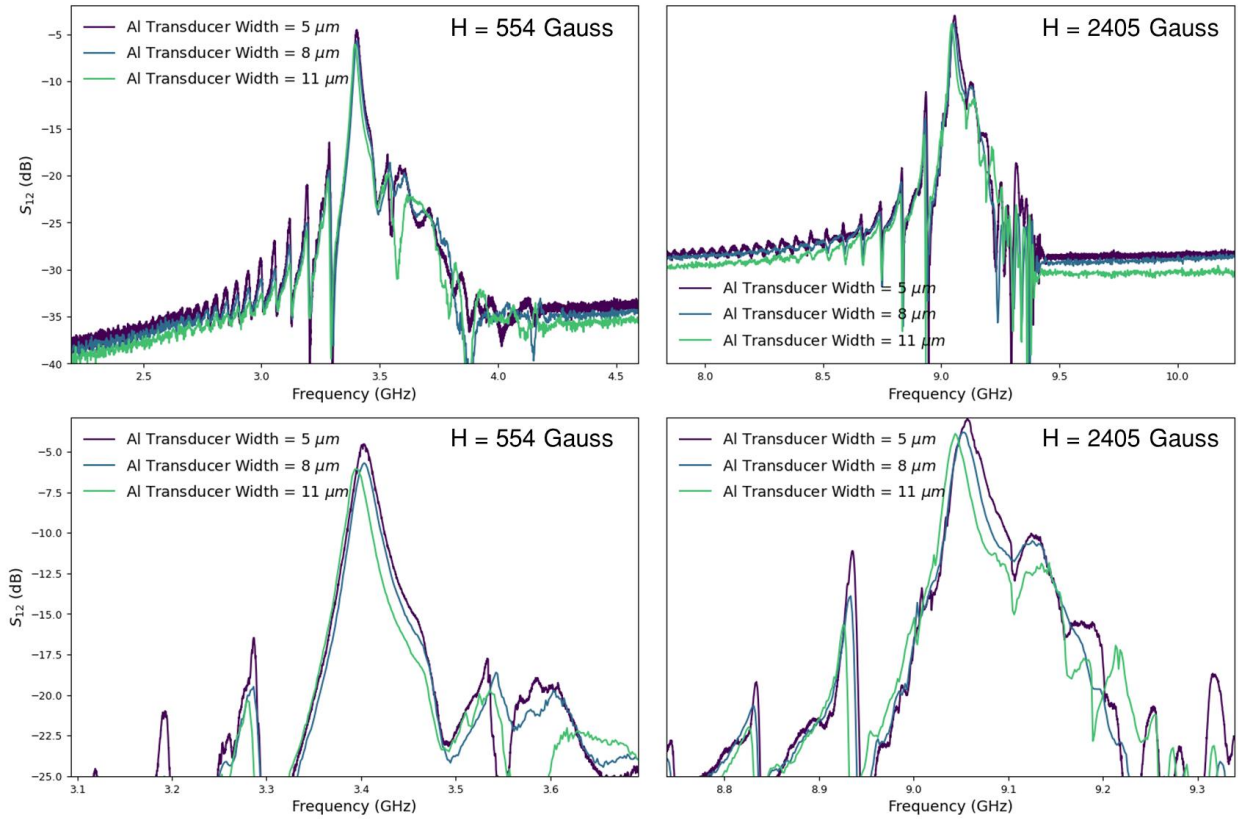

Supplementary Figure 9. Typical frequency responses of different MSSW filters with Al transducer widths of 5 μm, 8 μm, and 11 μm at a magnetic field of 554 and 2405 Gauss. The MSWF is designed with constant W = 200 μm and L = 70 μm.

**Supplementary Note 6: Length Effect of Magnetostatic Wave Resonators (MSWR)**

Supplementary Figure 10 compares the effect of length on MSWR $K^2$, Q-factor, FoM, and magnetostatic resistance $R_m$.

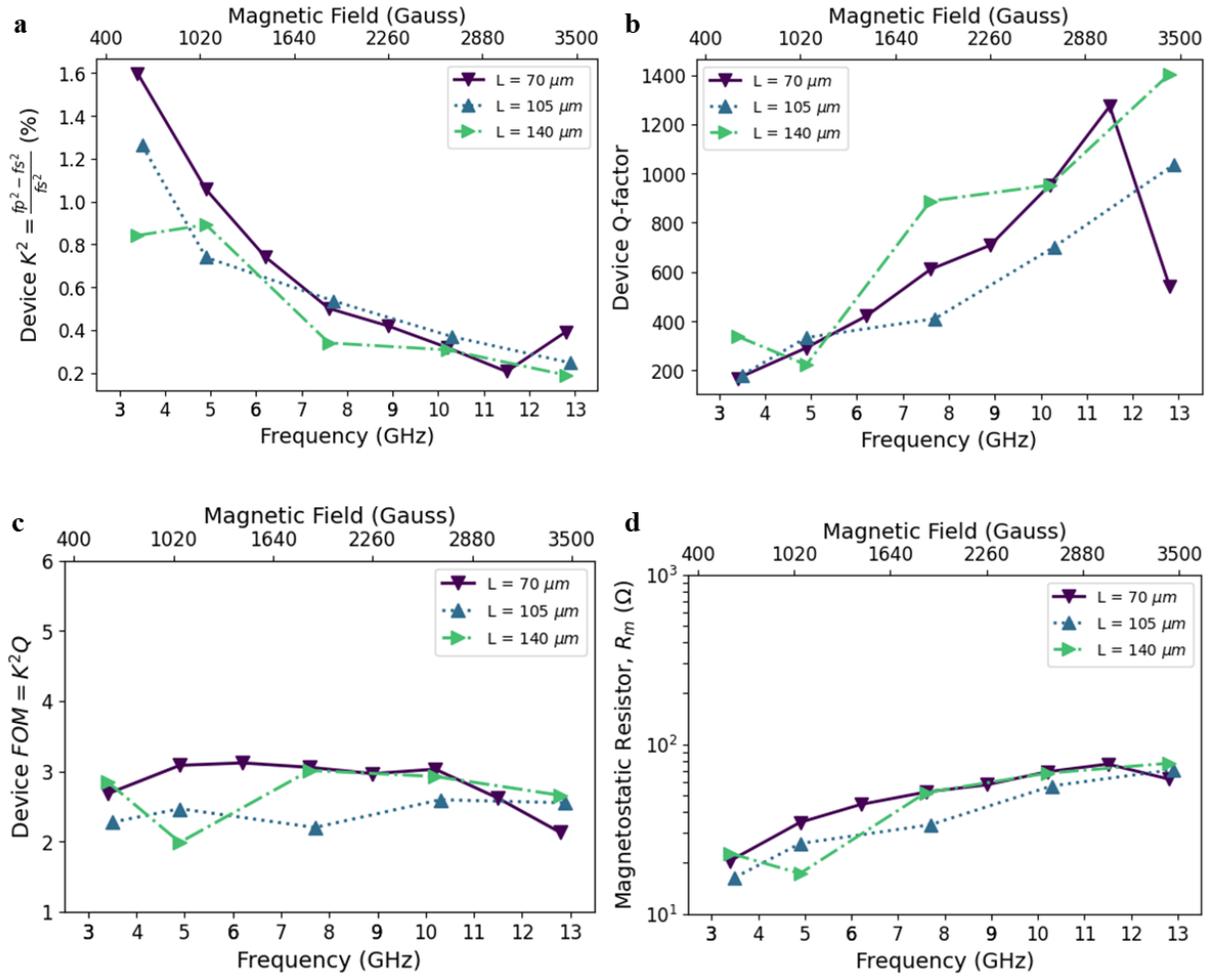

Supplementary Figure 10. Comparison of MSWR for various YIG cavity lengths with a constant width of 150 μm. (a) The effect of YIG cavity length on the device coupling coefficient, (b) The effect of YIG cavity length on the device Q-factor, (c) The effect of YIG cavity length on the device FoM, (d) The effect of YIG cavity length on the magnetostatic resistance, $R_m$.

**Supplementary Note 7: Impedance Matching of the Magnetostatic Wave Filters (MSWF)**

Figure 3 (d) illustrates that $R_m$ increases with the external magnetic field. For MSWR with W = 150 μm, $R_m$ is greater than 50 Ω for frequencies above 7.6 GHz and increases to 77 Ω at the frequency of 11.5 GHz. For the MSWR with W = 600 μm, the $R_m$ is greater than 50 Ω for all frequencies with a minimum of 115 Ω at 3.4 GHz and the maximum of 363 Ω at 11.6 GHz. This difference in $R_m$ causes different degrees of impedance mismatch for the MSWF vs. frequency.

Supplementary Figure 11 shows the return loss of the MSWF at the frequency where the $S_{12}$ reaches its peak. Due to the increase of the magnetostatic resistance of the MSWF with W = 600 μm, this return loss decreases from 18 dB and 22 dB at 3.4 and 6.3 GHz, respectively, to 14 dB and 8 dB at 9 GHz and 11.6 GHz, respectively. As a contrast, the MSWF with W = 150 μm shows a minimum return loss of 6 dB at 3.4 GHz and achieves its maximum of 13 dB at 10.2 GHz. For the MSWF with W = 150 μm or W = 200 μm, the maximum of $S_{11}$ is at the same frequency with the maximum of $S_{12}$, indicating the same fundamental mode. However, the large maximum $R_m$ of the devices with wider YIG cavities caused the maximum return loss to occur at a higher order mode of the MSWF where the $R_m$ of that higher order mode is closer to the 50 Ω termination impedance. At the fundamental mode where the maximum impedance is achieved, most of the signal is reflected back to the source and thus the insertion loss of the fundamental mode is high.

Supplementary Figure 12 depicts the insertion loss of the MSWF when terminated with 50 Ω. Due to the difference in the degree of the impedance matching, the insertion loss of the filters with a narrower YIG cavity (W = 150 or 200 μm) are more consistent across the frequency band whereas the insertion loss of the wide YIG cavity (W = 400 or 600 μm) decreases with increasing frequency.

To reduce the loss of the W = 600 μm devices across the filter band, an external series inductor and parallel capacitor can be used to achieve better matching to the 50 Ω termination impedance, but this requires matching networks that are tunable with frequency. Therefore, for the purpose of achieving a broad frequency tuning range of the filters, the MSWR should not only achieve the maximum FOM but also maintain good impedance matching to the 50 Ω source impedance across the tunable frequency range.

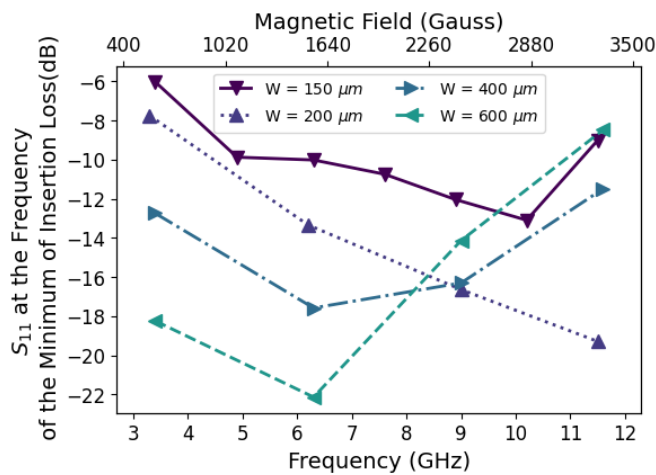

Supplementary Figure 11. $S_{11}$ where $S_{12}$ reaches the maximum vs. frequency.

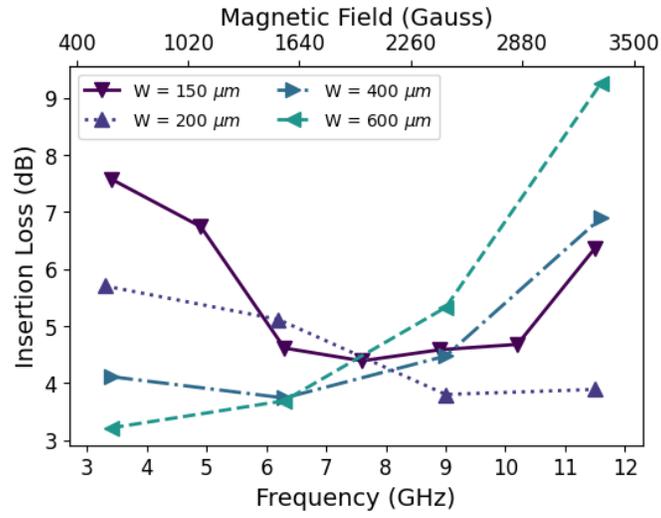

Supplementary Figure 12. Insertion loss vs. frequency. All the devices are made with aluminum transducers with width of 7 μm.

**Supplementary Note 8: Frequency Tunability of Magnetostatic Wave Filters (MSWF)**

Supplementary Figure 13 illustrates the tunability of the MSSW filters via applied magnetic field. In all the four different devices, the relationship of the main resonance frequency with respect to the applied magnetics bias field is linear with a slope of 2.9 MHz/Gauss. Here the resonance frequency is defined as the frequency where the loss reaches its minimum. This tunability is similar to devices reported in previous studies. [18, 19]

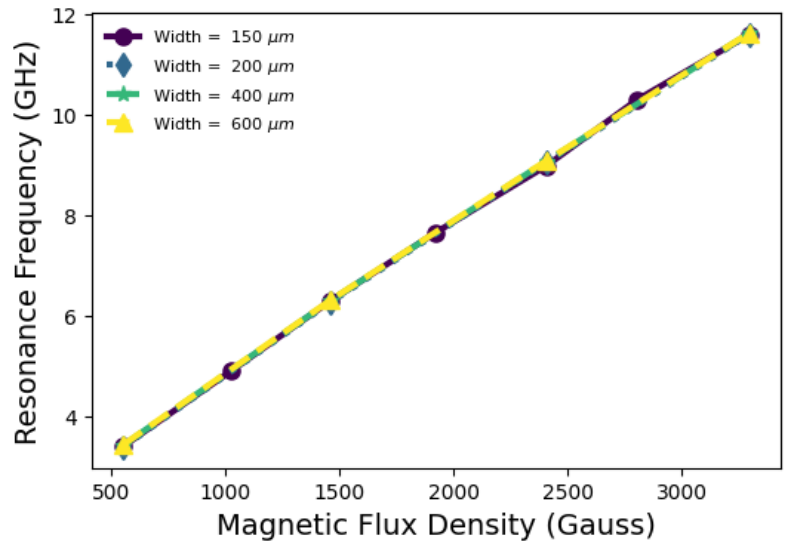

Supplementary Figure 13. Tunability of MSSW resonators. resonance frequency vs. applied magnetic field.

**Supplementary Note 9: Mode analysis of Magnetostatic Wave Resonators (MSWR)**

Because the MSSW travels along the surface of the YIG film and is reflected to the other side of the YIG surface at the straight edges, the resonant conditions are when the following equation is met:[20]

$$2k_y L = 2\pi n \qquad n = 1,2,3, \ldots \ldots \tag{S8}$$

Where $k_y$ is the average wavenumber for the top and bottom surfaces. As the length increases, the spacing between the resonance modes decreases in wavenumber, due to the strong dispersion of the MSSW. Although a wider MSWR can provide higher FoM and lower insertion loss, it also contains additional spurious responses. The dispersion relation with the width modes can be expressed as: [21]

$$\exp(2Md) = \frac{\Omega_m M + \Omega_k + (\Omega_H^2 - \Omega^2)(M - N)}{\Omega_H M - \Omega_k + (\Omega_H^2 - \Omega^2)(M + N)} \times \frac{\Omega_m M - \Omega_k + (\Omega_H^2 - \Omega^2)(M - N\tanh(Nt))}{\Omega_m M + \Omega_k + (\Omega_H^2 - \Omega^2)(M + N\tanh(Nt))} \tag{S9}$$

Inside the film, we have

$$M^2 = k_x^2 = \frac{(\frac{n\pi}{W})^2}{\mu_1} + k_y^2 \tag{S10}$$

And outside

$$N^2 = k_x^2 = (\frac{n\pi}{W})^2 + k_y^2 \tag{S11}$$

where $\mu_1 = 1 - \Omega_H/(\Omega^2 - \Omega_H^2)$, $\Omega = \omega/\nu 4\pi M_s$, $\Omega_H = H/4\pi M_s$, $\omega$ is the angular frequency, $\nu$ is the gyromagnetic ratio of 2.8 MHz/Oe for YIG, and $4\pi$ Ms is the saturation magnetization of the YIG film which is 1780 Gauss, H is the external magnetic field applied, t is the distance between YIG film and ground plane which can be chosen as any arbitrary large number here.

The calculated results are shown in Supplementary Figure 14 and the measurements of the $S_{12}$ frequency response for the MSWFs with widths of 200 μm, 400 μm, and 600 μm are shown in Supplementary Figure 15. Because the width of the MSWF is much smaller than the electromagnetic wavelength at these frequencies and the current remains almost constant along the transducer, only odd order width modes can be excited. The frequency spacing between two adjacent width modes increases for narrower YIG cavities. The frequency response of W = 600 μm and W = 150 μm confirms this result. There is also a frequency shift for the main resonance mode with increases in width, as can be seen from the dispersion curve where a constant wavenumber corresponds to a higher frequency as the width increases. As both theoretical calculation and experimental measurements confirm, the MSSW becomes more dispersive at higher frequency for wider MSSW filters.

Supplementary Figure 16 and 17 illustrate the $S_{11}$ frequency response of devices with varying widths and lengths. Consistent with the earlier findings, it can be observed that longer and wider MSWFs exhibit more spurious responses. Additionally, wider MSWFs display a lower minimum value of $S_{11}$. However, the length of the MSWF does not significantly affect the minimum value of $S_{11}$, as the increase in $R_m$ is primarily associated with the width rather than the length. These observations further support the conclusions regarding the influence of width and length on the spurious responses and impedance characteristics of the MSWRs.

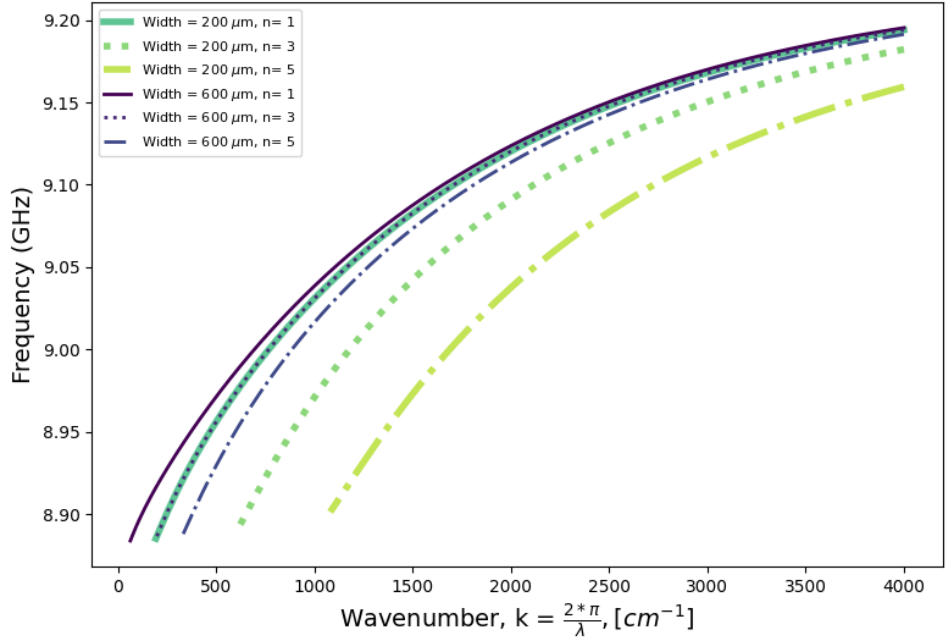

Supplementary Figure 14. Dispersion curves for MSWF width modes with magnetic flux density of 2405 Gauss.

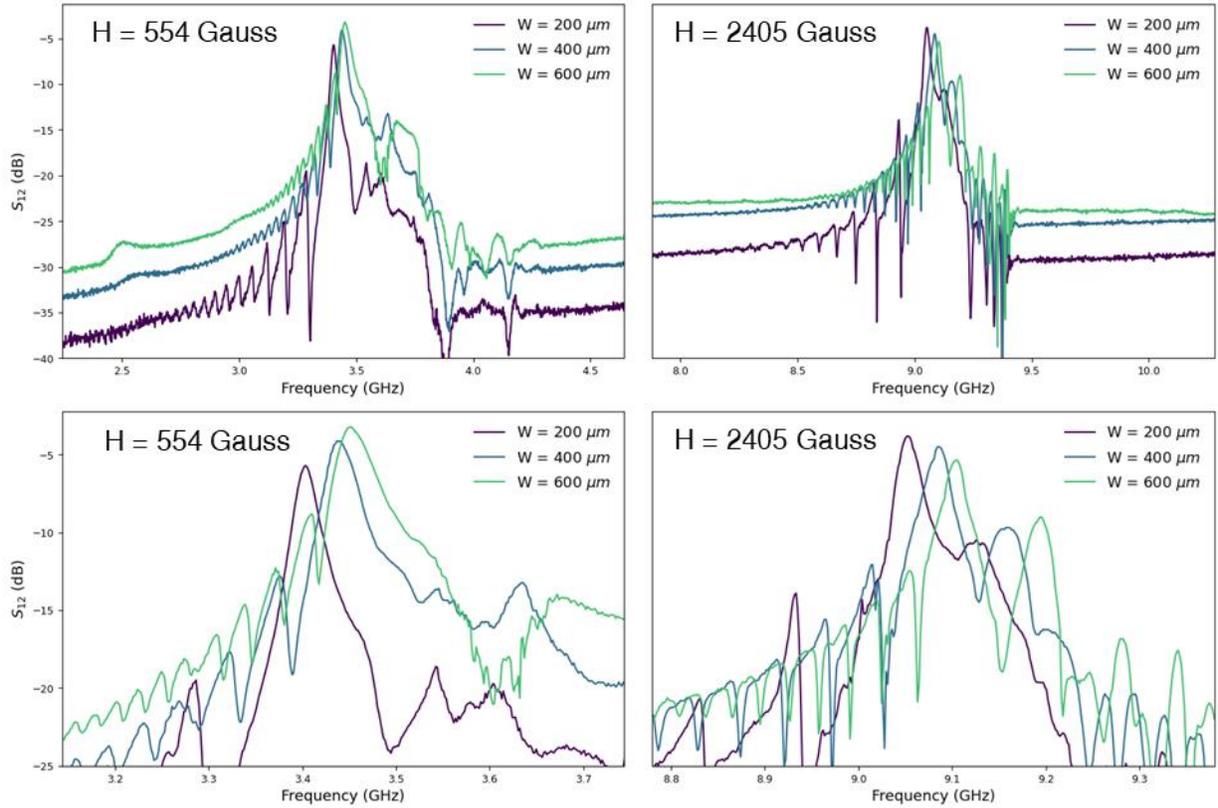

Supplementary Figure 15. $S_{12}$ frequency response of the MSWF at different magnetic flux density (H). The impact of YIG cavity width on the frequency response with constant length of 70 μm.

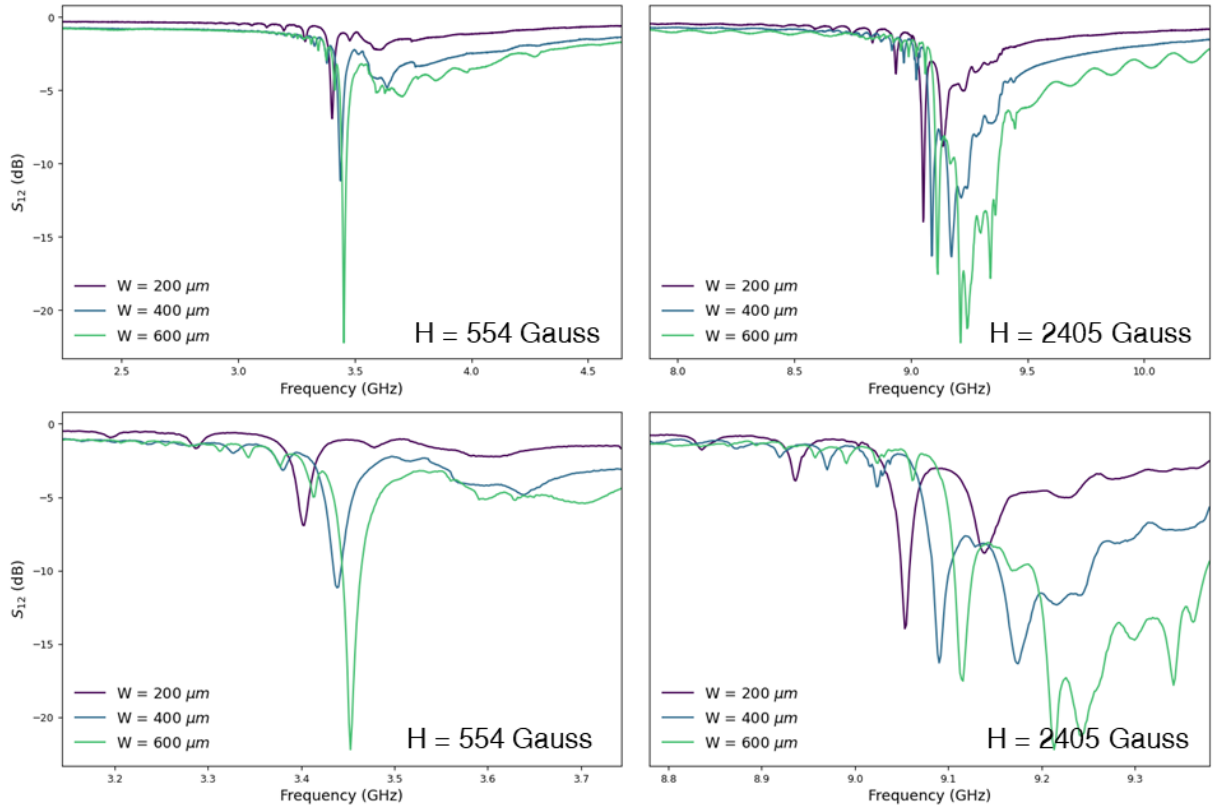

Supplementary Figure 16. $S_{11}$ frequency response of the MSWF at different magnetic flux density (H). The impact of YIG cavity width on the frequency response with constant length of 70 μm.

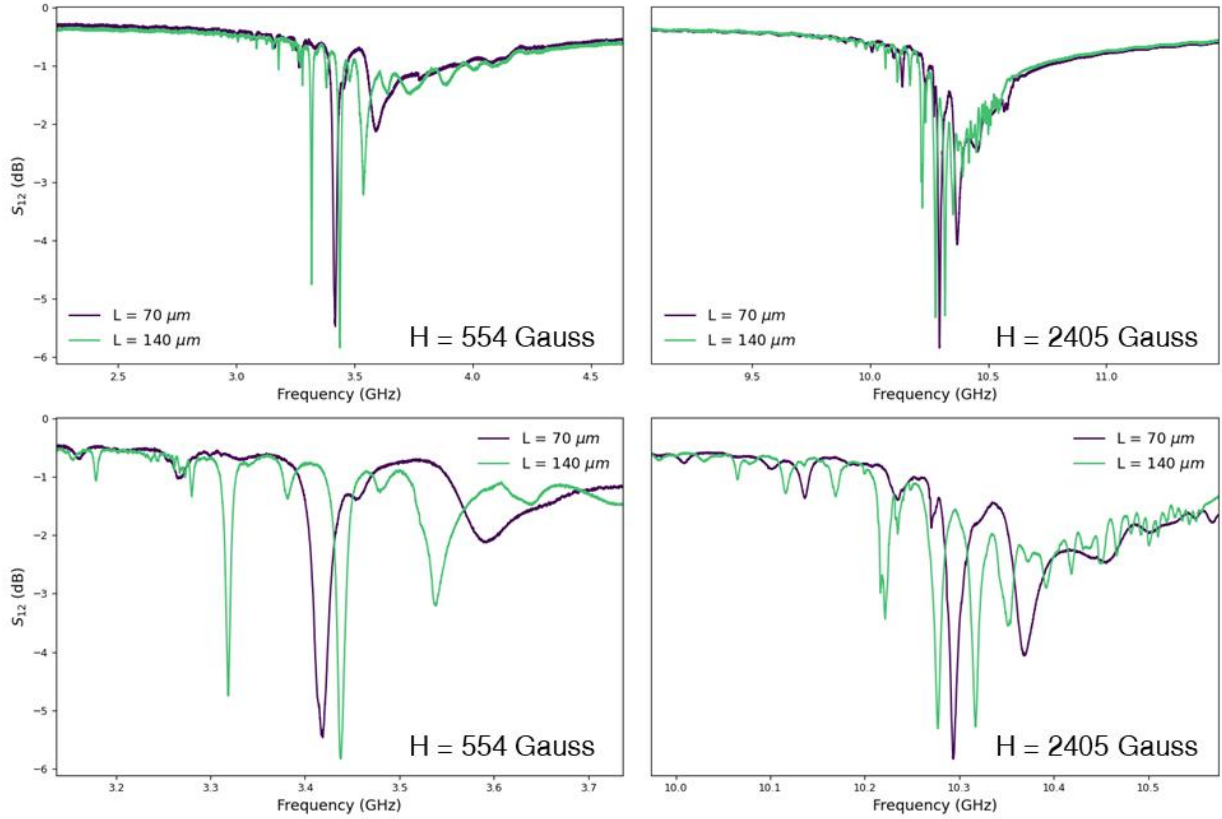

Supplementary Figure 17. S$_{11}$ frequency response of the MSWF at different magnetic flux density (H). The impact of YIG cavity length on the frequency response with constant length of 150 μm.

**Supplementary Note 10: Circuit Modeling Procedure for Magnetostatic Wave Filters (MSWF)**

Using the prescribed circuit model procedure, it is possible to accurately fit the frequency response of the two-port MSWF. By applying the appropriate circuit parameters and modeling techniques, the circuit model can replicate the observed frequency characteristics of the MSWF. This allows for a better understanding and analysis of the filter's performance and behavior. For the MSWF, the off-resonance $Z_{11}$ impedance is the same as the $Z_{series}$, and the off-resonance $Z_{12}$ impedance is mainly due to the impedance of the inductor $L_c$, which can be calculated by the imaginary part of the off-resonance $Z_{12}$. Thus, the $L'_s$ in MSWF can be expressed as:

$$L'_s = L_{s1} - L_c \tag{S12}$$

Supplementary Figure 18 shows the comparison of the $Z_{12}$ and $S_{12}$ measurements and the single mode circuit model. In this single-mode circuit model, only one resonance tank was used. Its parameters can be found in Supplementary Table 1. The $L'_s$ and $L_c$ of the circuit is 0.34 nH and 30.64 pH, respectively.

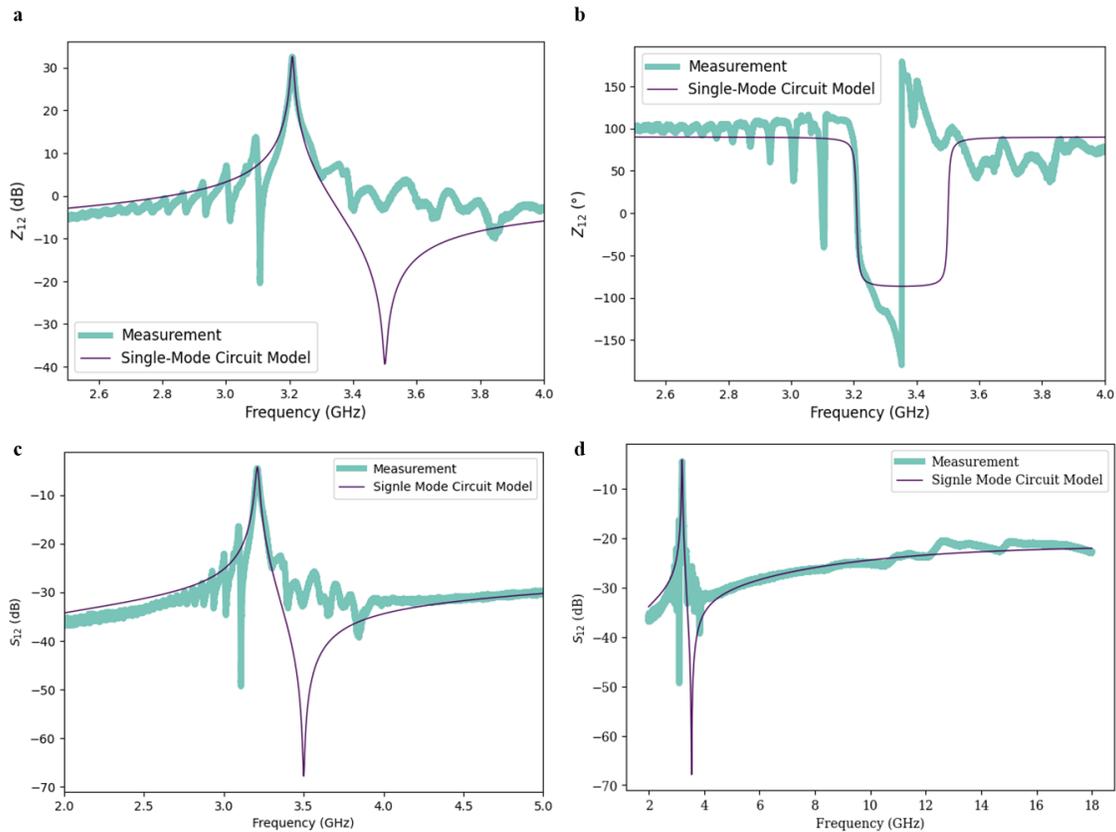

Supplementary Figure 18. Examples of the single-resonance recursive fitting using measurement of MSWF with width of 200 μm and length of 70 μm with transducers width of 4 μm. (a) Magnitude of $Z_{12}$, (b) Phase of $Z_{12}$, (c) Magnitude of $S_{12}$ with frequency from 2 to 5 GHz, (d) Magnitude of $S_{12}$ with frequency from 2 to 16 GHz.

**Supplementary Note 11: 1 dB Compression Measurement**

High power measurements of the fabricated filter were carried out on a MSSW filter with W = 150 μm and L = 70 μm. Supplementary Figure 19 shows the frequency response of $S_{12}$ with different input power. Supplementary Figure 20 reports the input 1dB compression point with respect to different center frequencies.

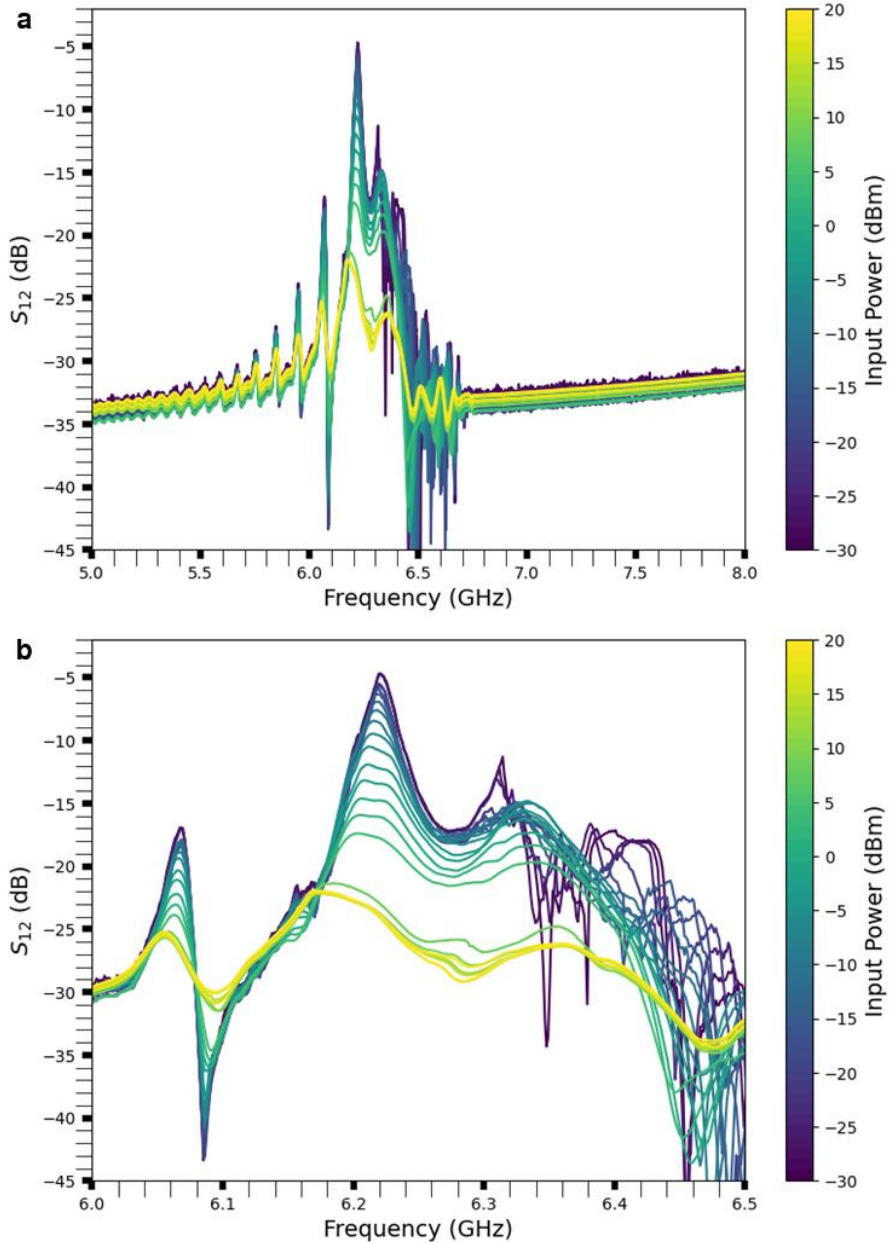

Supplementary Figure 19. Comparison of S12 with different input power by using width = 150 μm, length = 70 μm with transducers on top of the YIG devices. The input power increased from -30 dBm to 20 dBm with a step of 2 dBm.

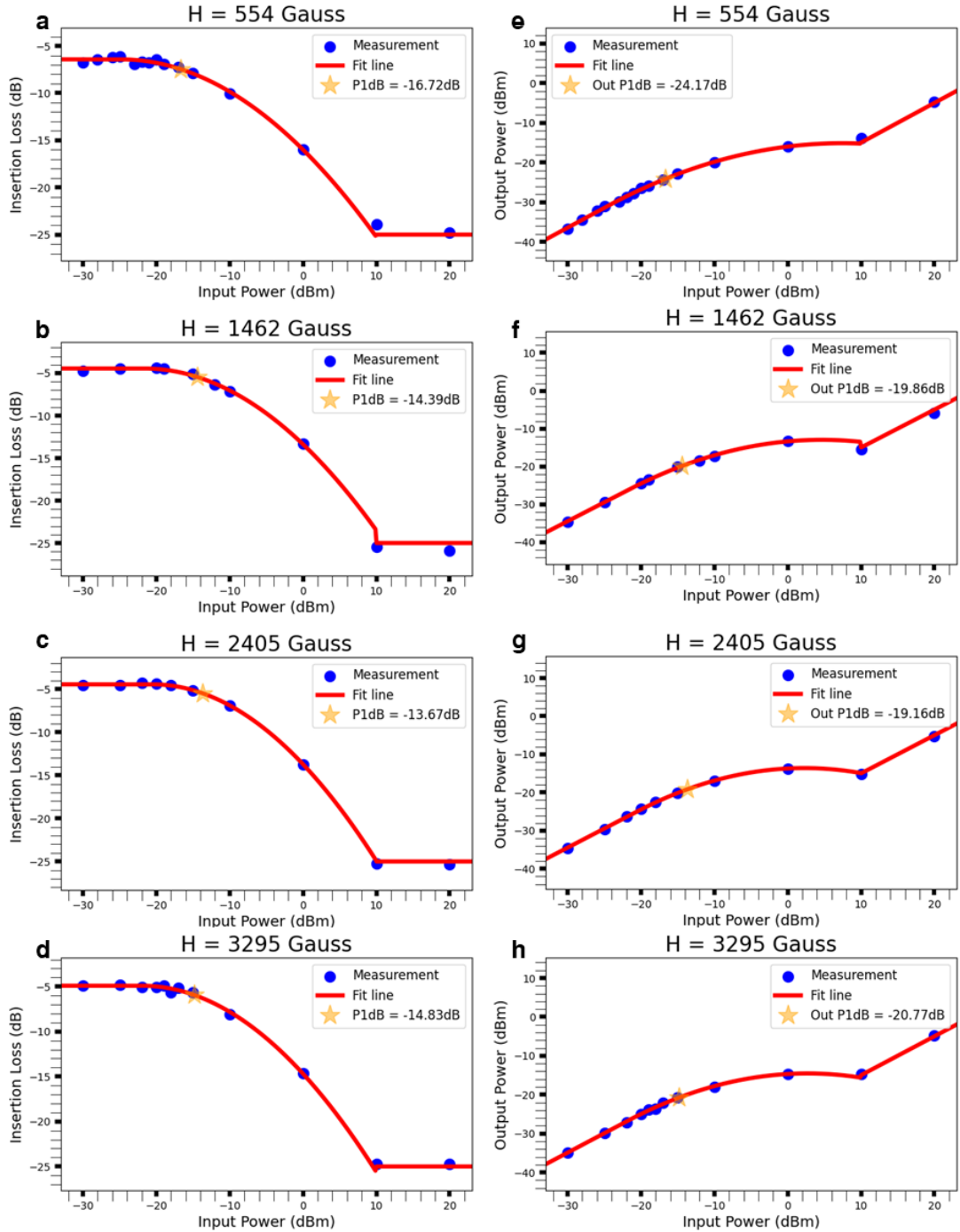

Supplementary Figure 20. Comparison of insertion loss and output power with different input powers for the MSWF with width = 150 μm, length = 70 μm. Effect on insertion loss with (a) frequency of 3.4 GHz, (b) frequency of 6.2 GHz, (c) frequency of 8.9 GHz, (d) frequency of 11.6 GHz. Effect on output power with (e) frequency of 3.4 GHz, (f) frequency of 6.2 GHz, (g) frequency of 10.3 GHz, (h) frequency of 11.6 GHz.

**Supplementary Note 12: Intermodulation intercept point (IIP3) Measurement**

Supplementary Figure 21 shows the schematic for the IIP3 measurement setup. For IIP3 measurement, the two-tone signals ($f_1$ and $f_2$) were generated using two signal generators (HP 83712A and Gigatronics 2520A). The signal was amplified by a Minicircuit ZX60-83LN-S+ or Minicircuit ZX60-183-S+ for out-of-band IIP3 measurement. For in-band IIP3 measurement, two 30 dB attenuators (BW-S30W2+) were used. Two bandpass filters (Minicircuit ZBSS-10G-S+ for frequencies above 8 GHz, and ZBSS-6G-S+ for frequencies below 8 GHz) and two circulators (DiTom Microwave DMC6018 for frequencies above 6 GHz, and DiTom Microwave D3C2060 for frequencies below 6 GHz) were utilized to clean the signal from the amplifiers. After that, two signals were combined in a power combiner (Krytar 6020265) and connected to another circulator and finally to the DUT (the MSWF device under test). The output of the YIG filter device was connected to another circulator and then to the spectrum analyzer (HP 8563E).

SI Figure 22, 23, 24, 25 and 26 show the in-band and out-of-band IIP3 measurements.

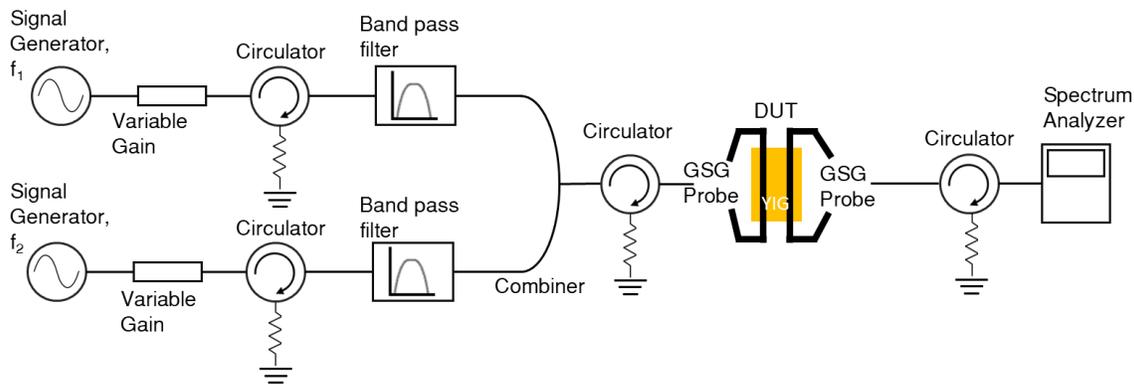

Supplementary Figure 21. IIP3 measurement setup schematic.

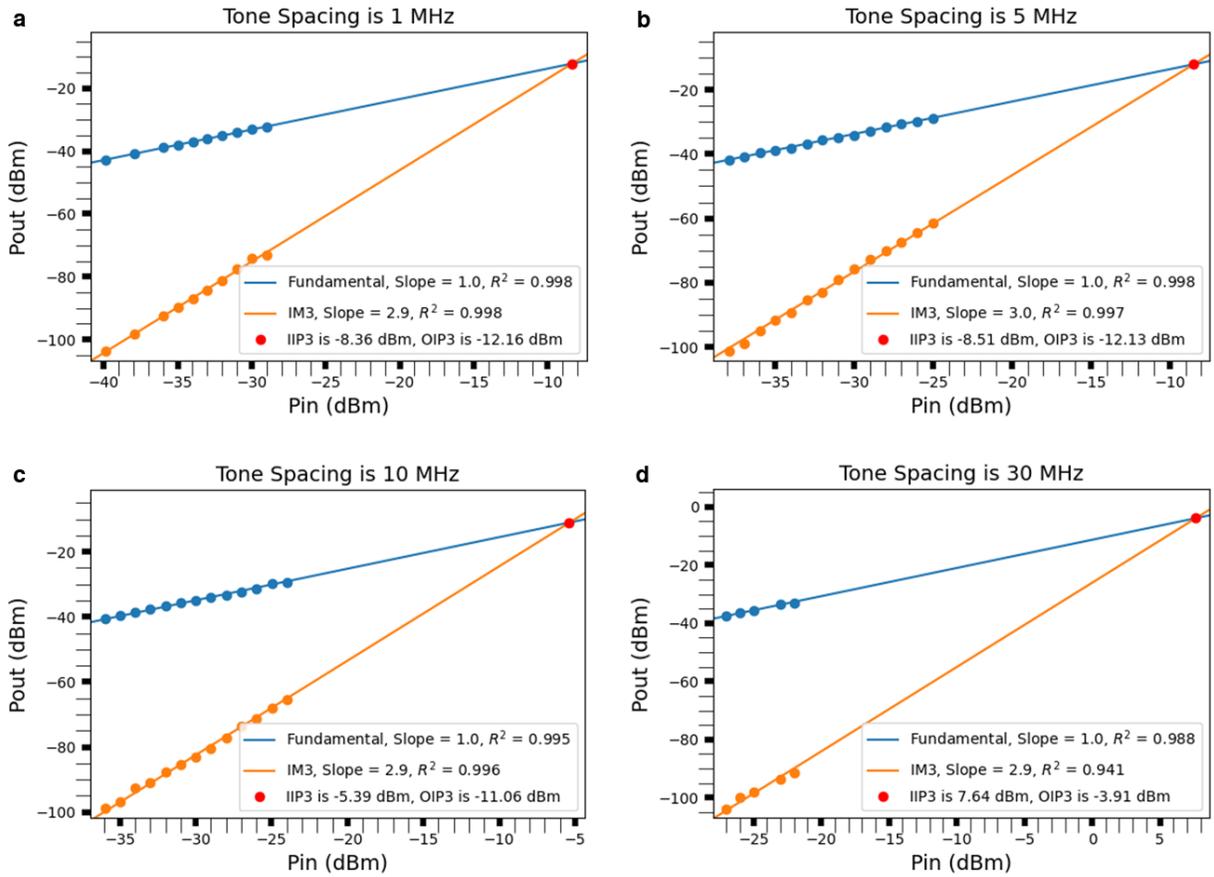

Supplementary Figure 22. In-band third order input intercept point (IIP3). The device is measured at a frequency of 7.6 GHz with tone spacings of (a) 1 MHz, (b) 5 MHz, (c) 10 MHz, and (d) 30 MHz.

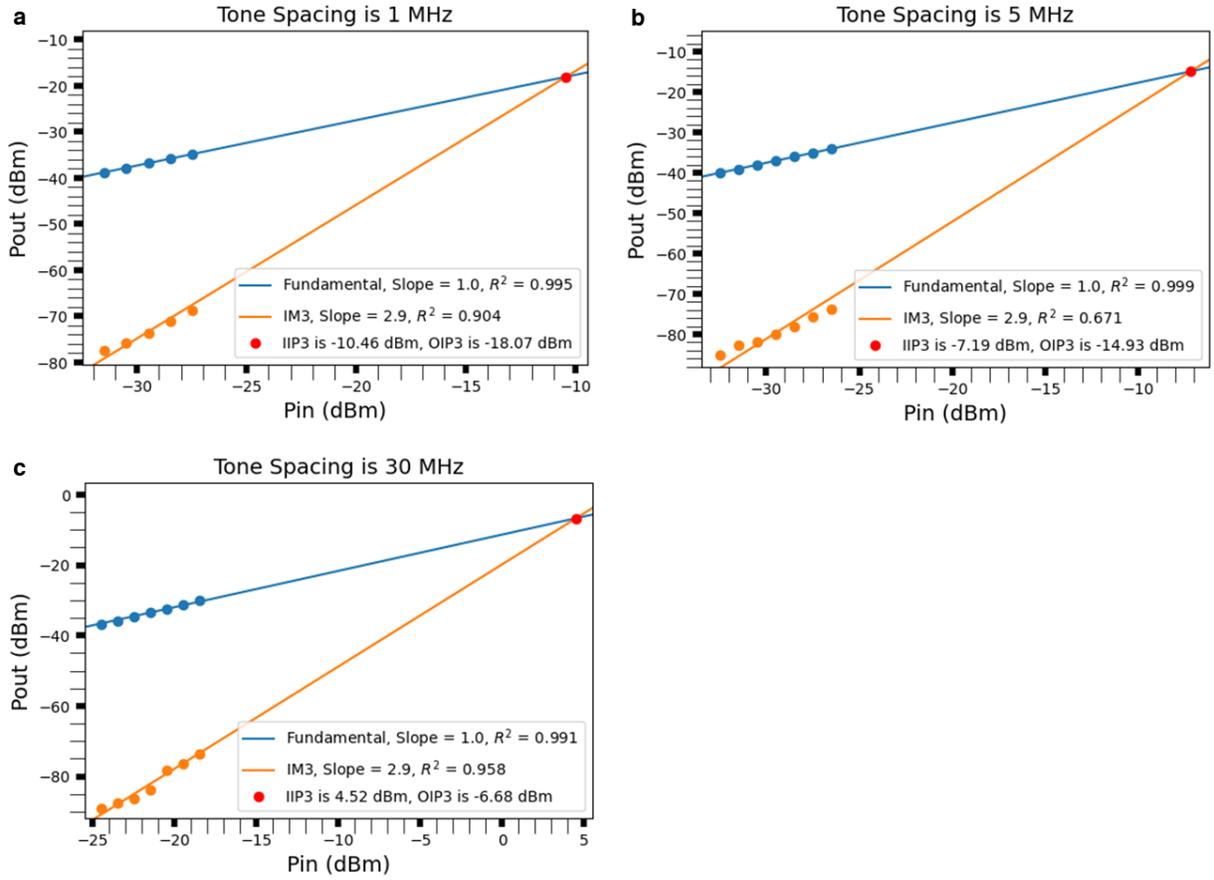

Supplementary Figure 23. In-band third order input intercept point (IIP3). The device is measured at a frequency of 10.1 GHz with tone spacings of (a) 1 MHz, (b) 5 MHz and (c) 30 MHz.

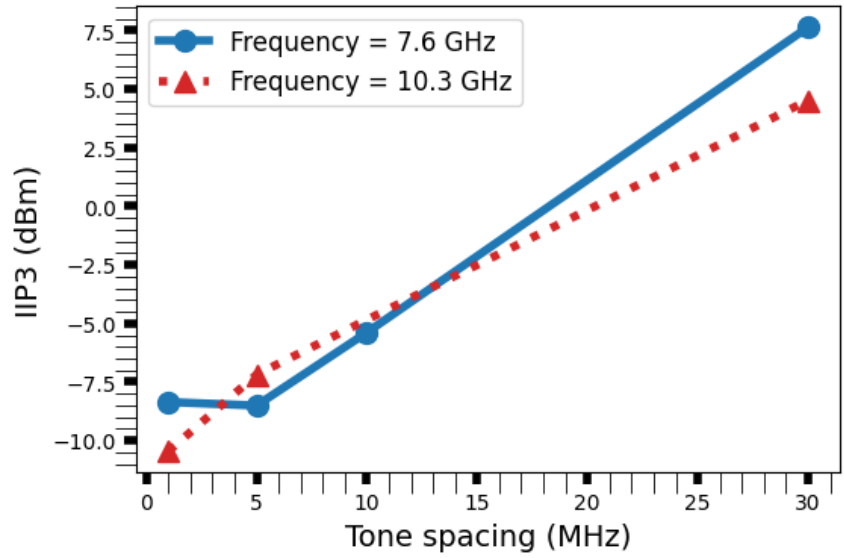

Supplementary Figure 24. A summary of the in-band third order input intercept point (IIP3) measurement result.

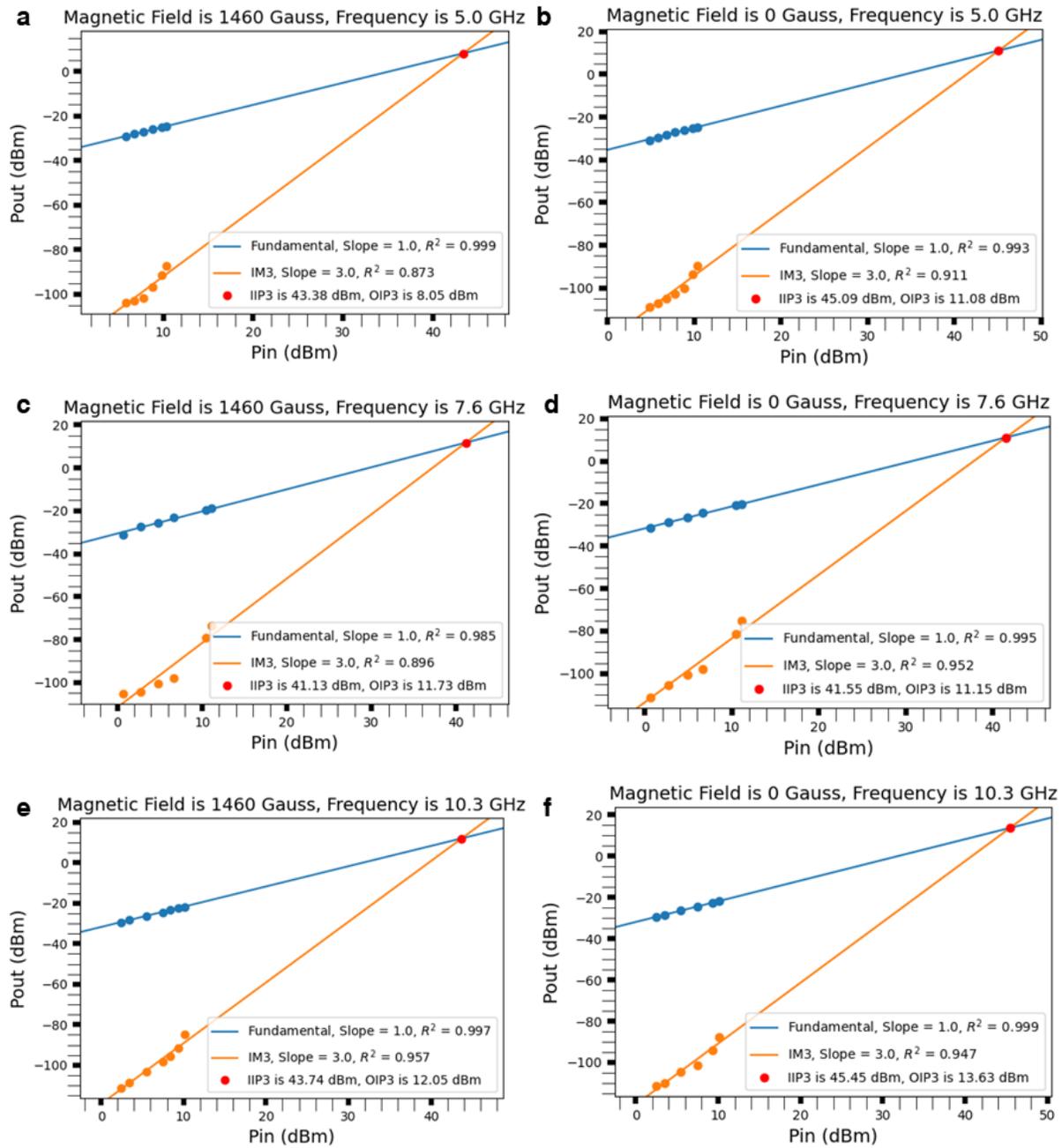

Supplementary Figure 25. Out-of-band (OOB) IIP3 measurement by using width = 150 μm, length = 70 μm devices. The device is measured at frequencies of 5.0 GHz (a) and (b), 7.6 GHz (c) and (d), 10.3 GHz (e) and (f), with an applied magnetic field of 1460 Gauss (a), (c), and (e) and without magnetic field (b), (d), and (f). The tone spacing is 50 MHz.

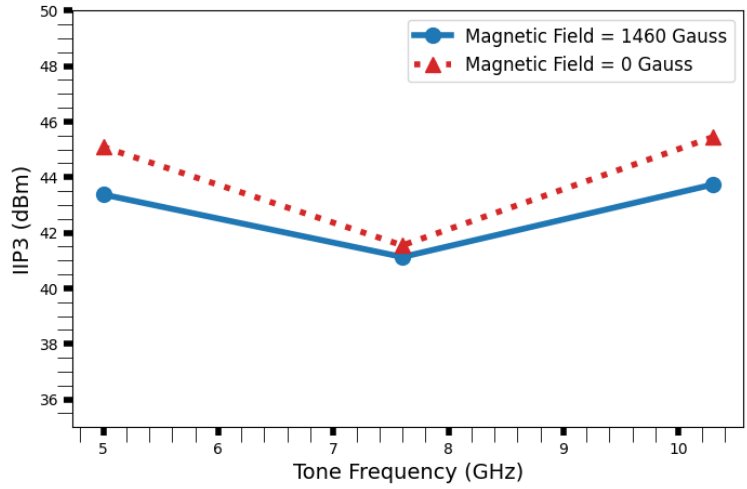

Supplementary Figure 26. Summary of out-of-band (OOB) IIP3 measurement by using width = 150 μm, length = 70 μm devices.

**Supplementary Note 13: Capacitor Voltage and Coil Current During Capacitor Discharging**

To generate a pulse of current for magnetizing/demagnetizing the AlNiCo magnets, a capacitor was used. First, the capacitor was charged to a voltage chosen to produce the desired magnetic field. Subsequently, it was connected to the coil and discharged, which produced a current response according to a series RLC circuit. Supplementary Figure 27 shows the measured voltage across the capacitor and current flowing through the coil during the capacitor discharging. The capacitor used in the experiment had a capacitance of 270 μF and was charged to 19 V. The peak current flowing through the coil was approximately 27 A. The amplitude of the current pulse is proportional to the capacitor charging voltage. By controlling the capacitor charging voltage and the current pulse amplitude, the remanent flux of the AlNiCo magnets can be adjusted. Therefore, tuning of the magnetic field at the YIG chip is realized.

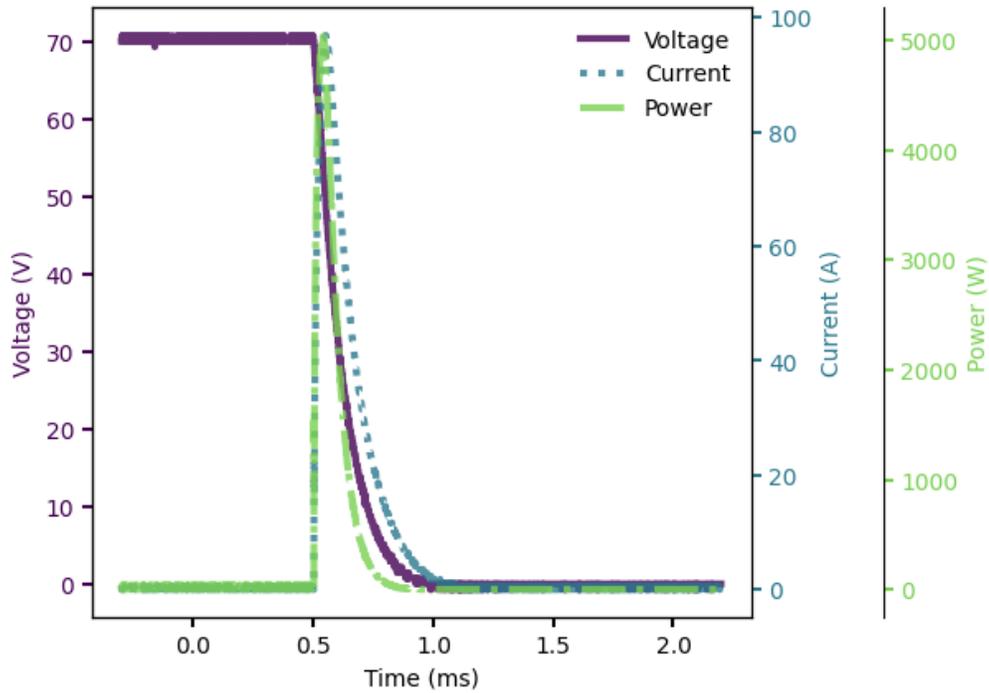

Supplementary Figure 27. Capacitor voltage and coil current during capacitor discharging.

**Supplementary Note 14: Magnetic Feld Uniformity of the Magnetic Biasing Circuit**

Instead of using a 2 mm thick yoke, a thinner 0.5 mm thick yoke was initially explored to create a separate magnetic biasing circuit. Supplementary Figure 28 shows the measured magnetic flux density as a function vertical position for different capacitor charging voltages. In contrast to the flat flux vs. vertical position observed in Fig. 5 (a), the magnetic flux density profile of the circuit with the 0.5 mm thick yoke exhibits a sharp peak at the center. This indicates that the magnetic biasing circuit is not as uniform as the 2 mm thick yoke. Supplementary Figure 29 shows one example of the filter frequency response where the MSWF was integrated with the magnetic biasing circuit with 0.5 mm thick yoke. In this device, due to the nonuniform magnetic field that is applied to the YIG, the velocity is different in different regions of the MSWF. Thus, return loss reduces and the insertion loss increases. Due to the almost constant bandwidth of the MSSW filter, the requirement of magnetic field uniformity is an absolute constant irrespective of the frequency and the frequency tuning ratio of the MSW device is about 2.9 MHz/Gauss. Thus, it is important for the magnetic bias circuit to provide a high uniformity at all the magnetic bias field levels.

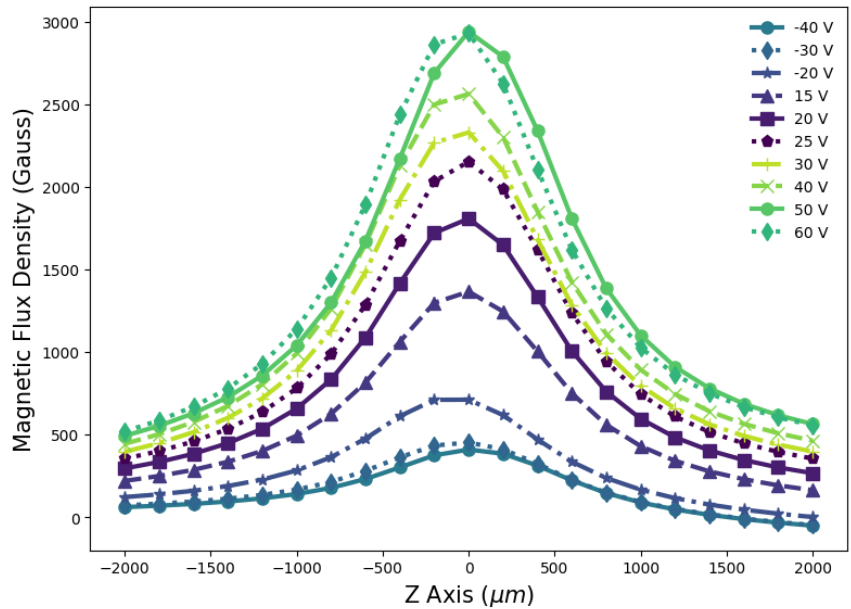

Supplementary Figure 28. Measured magnetic flux density of the magnetic biasing circuit with thin yoke under different capacitor charging voltages.

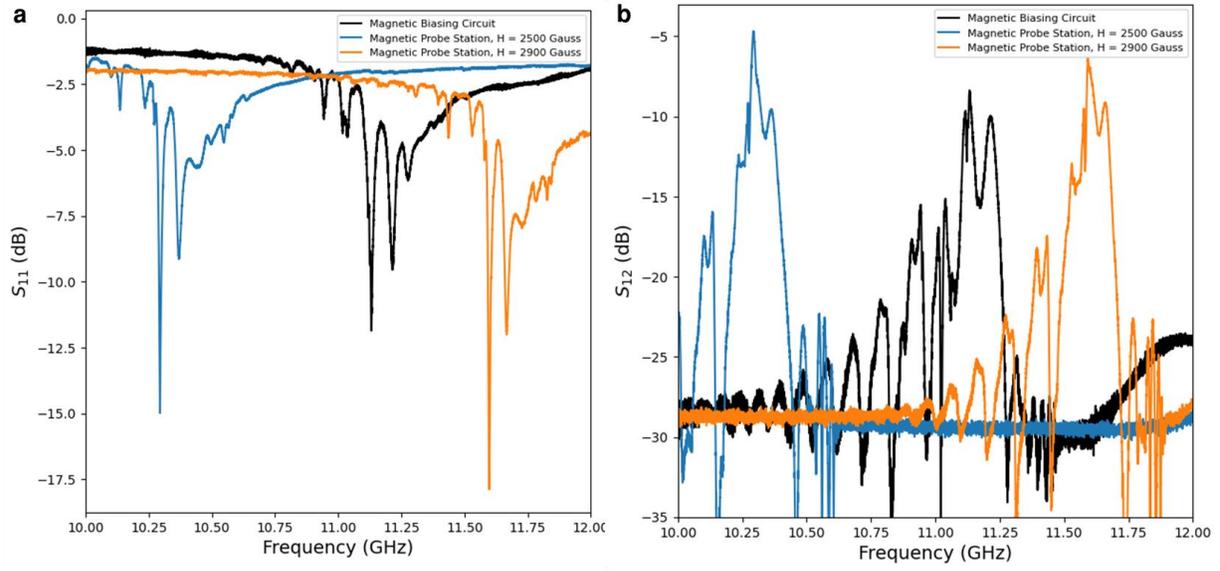

Supplementary Figure 29. Comparison of the frequency response of the MSWF inside the magnetic bias circuit and magnetic probe station where the applied magnetic field of the magnetic bias circuit is nonuniform. (a) $S_{11}$ frequency response, (b) $S_{12}$ frequency response of the width = 150 μm, length = 70 μm MSWF.

**Supplementary Note 15: Magnetically Tunable Notch Filter**

Supplementary Figure 30 illustrates an MSWF reconfigured into a band stop filter, where the two ports were connected by aluminum transducers routing on top of the YIG cavity. To minimize inductance and resistance of the line, the aluminum transducers were designed to be as wide as possible in regions not on top of the YIG. Each YIG cavity had a width of 200 μm and a length of 70 μm. In Supplementary Figure 31, the measured $S_{12}$ frequency response of the band stop filter comprising 9 stages of YIG cavity. This filter demonstrates less than 1.5 dB of pass band insertion loss and greater than 20 dB rejection in the tunable notch.

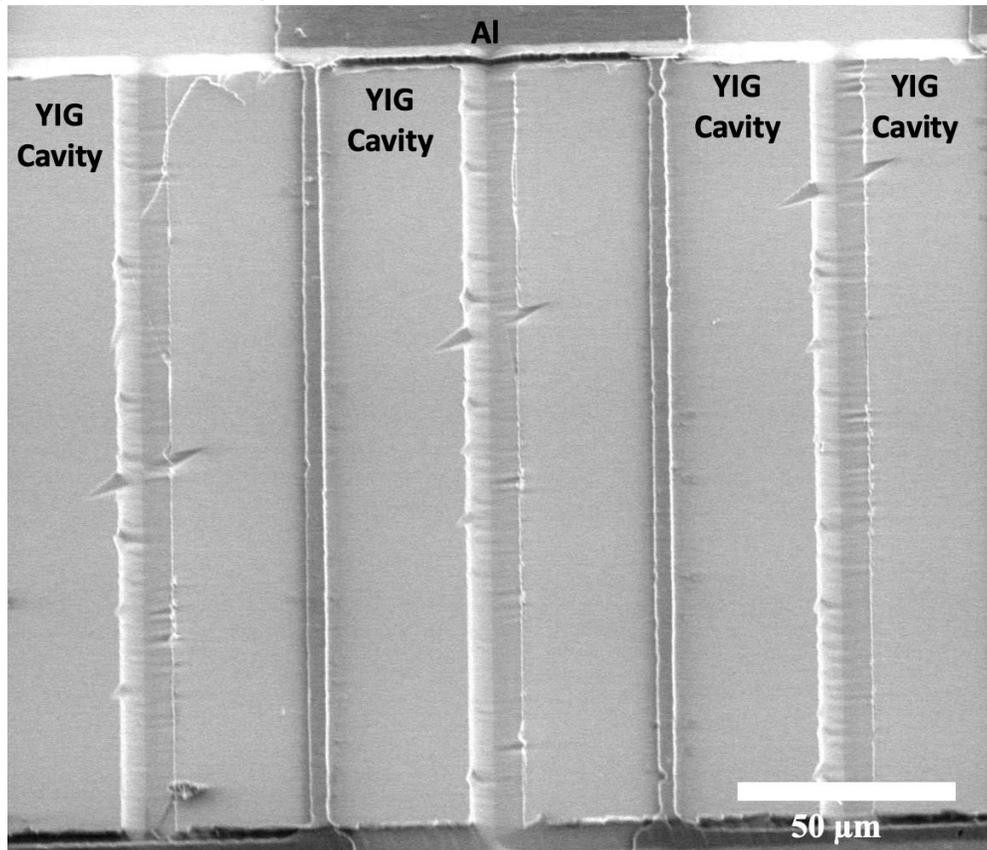

Supplementary Figure 30. SEM image of magnetostatic wave notch filter.

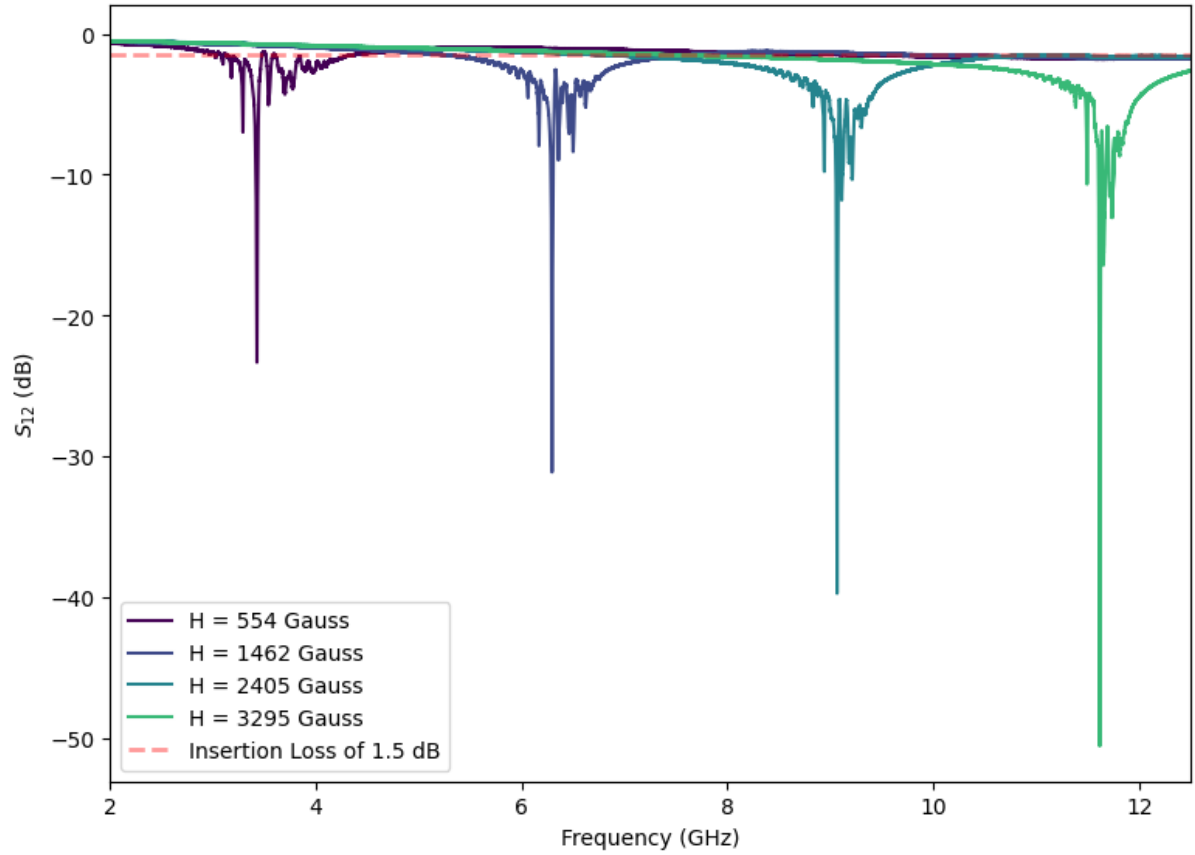

Supplementary Figure 31. Measured $S_{12}$ frequency response of magnetostatic wave notch filter measured with magnetic field supplied by the magnetic probe station.

**Reference**

1. Castéra JP, Hartemann P. Magnetostatic wave resonators and oscillators. *Circuits, Systems and Signal Processing* **4**, 181-200 (1985).

2. Tsai CS, Qiu G. Wideband Microwave Filters Using Ferromagnetic Resonance Tuning in Flip-Chip YIG-GaAs Layer Structures. *IEEE Transactions on Magnetics* **45**, 656-660 (2009).

3. Wu J, Yang X, Beguhn S, Lou J, Sun NX. Nonreciprocal tunable low-loss bandpass filters with ultra-wideband isolation based on magnetostatic surface wave. *IEEE transactions on microwave theory and techniques* **60**, 3959-3968 (2012).

4. Du S, Yang Q-H, Fan X, Zhang H. High selectivity and compact tunable bandpass filter using YIG material. *Journal of Applied Physics* **133**,  (2023).

5. Micro Lambda Wireless Inc. MLFP 4 Stage Filter Data Sheet. *MLFP-42018*.

6. Chiou Y-C, Rebeiz GM. Tunable 1.55-2.1 GHz 4-Pole Elliptic Bandpass Filter With Bandwidth Control and > 50 dB Rejection for Wireless Systems. *IEEE Transactions on Microwave Theory and Techniques* **61**, 117-124 (2012).

7. MACOM Corporation. MA46 Series Surface Mount GaAs Tuning Varactors 0.75, 1.25, & 1.5 Gamma Hyperabrupt.

8. Wei Z, Yang T, Chi PL, Zhang X, Xu R. A 10.23–15.7-GHz Varactor-Tuned Microstrip Bandpass Filter With Highly Flexible Reconfigurability. *IEEE Transactions on Microwave Theory and Techniques* **69**, 4499-4509 (2021).

9. MACOM Corporation. MAVR-011020-1141 Solderable GaAs Constant Gamma Flip-Chip Varactor Diode

10. Courreges S, Li Y, Zhao Z, Choi K, Hunt A, Papapolymerou J. A Low Loss X-Band Quasi-Elliptic Ferroelectric Tunable Filter. *IEEE Microwave and Wireless Components Letters* **19**, 203-205 (2009).

11. Schuster C, *et al.* Performance analysis of reconfigurable bandpass filters with continuously tunable center frequency and bandwidth. *IEEE Transactions on Microwave Theory and Techniques* **65**, 4572-4583 (2017).

12. Nath J, *et al.* An electronically tunable microstrip bandpass filter using thin-film Barium-Strontium-Titanate (BST) varactors. *IEEE transactions on microwave theory and techniques* **53**, 2707-2712 (2005).

13. Sinanis MD, Adhikari P, Jones TR, Abdelfattah M, Peroulis D. High-Q High Power Tunable Filters Manufactured With Injection Molding Technology. *IEEE Access* **10**, 19643-19653 (2022).

14. Liu X, Katehi LPB, Chappell WJ, Peroulis D. High-Q Tunable Microwave Cavity Resonators and Filters Using SOI-Based RF MEMS Tuners. *Journal of Microelectromechanical Systems* **19**, 774-784 (2010).

15. Teledyne RF & Microwave Solutions. Teledyne RF&M YIG Supernotch Filters Overview Web. *https://wwwteledynedefenseelectronicscom/rfµwave/resources/Documents/Brochures/Teledyne%20RF&M%20YIG%20Supernotch%20Filters%20Overview%20Webpdf*.

16. Lu R, Li MH, Yang Y, Manzaneque T, Gong S. Accurate Extraction of Large Electromechanical Coupling in Piezoelectric MEMS Resonators. *Journal of Microelectromechanical Systems* **28**, 209-218 (2019).

17. COMSOL multiphysics® v. 6.6

18. Brown JA, Barth S, Smyth BP, Iyer AK. Compact Mechanically Tunable Microstrip Bandstop Filter With Constant Absolute Bandwidth Using an Embedded Metamaterial-Based EBG. *IEEE Transactions on Microwave Theory and Techniques* **68**, 4369-4380 (2020).

19. Dai S, Bhave SA, Wang R. Octave-Tunable Magnetostatic Wave YIG Resonators on a Chip. *IEEE Transactions on Ultrasonics, Ferroelectrics, and Frequency Control* **67**, 2454-2460 (2020).

20. Huijer E, Ishak W. MSSW resonators with straight edge reflectors. *IEEE Transactions on Magnetics* **20**, 1232-1234 (1984).

21. O'Keeffe TW, Patterson RW. Magnetostatic surface-wave propagation in finite samples. *Journal of Applied Physics* **49**, 4886-4895 (1978).